# IDENTIFICATION OF THE TRUE ELASTIC MODULUS OF HIGH DENSITY POLYETHYLENE FROM TENSILE TESTS USING AN APPROPRIATE REDUCED MODEL OF THE ELASTOVISCOPLASTIC BEHAVIOR

**A. Blaise*, S. André*, P. Delobelle**, Y.Meshaka***, C. Cunat***

** LEMTA-Nancy University-CNRS, 2 avenue de la Forêt de Haye, 54504, Vandoeuvre-Lès-Nancy, France*

*** LMA, Femto-ST, UFR Sciences, 24 chemin de l'Epitaphe, 25000, Besançon, France*

**** Institut Jean Lamour-Nancy University-CNRS, Parc de Saurupt, 54042 Nancy Cedex, France*

*E-mail: stephane.andre@ensem.inpl-nancy.fr*

Abstract

The rheological parameters of materials are determined in the industry according to international standards established generally on the basis of widespread techniques and robust methods of estimation. Concerning solid polymers and the determination of Young's modulus in tensile tests, ISO 527-1 or ASTM D638 standards rely on protocols with poor scientific content: the determination of the slope of conventionally defined straight lines fitted to stress-strain curves in a given range of elongations. This paper describes the approach allowing for a correct measurement of the instantaneous elastic modulus of polymers in a tensile test. It is based on the use of an appropriate reduced model to describe the behavior of the material. The model comes a thermodynamical framework and allows to reproduce the behavior of an HDPE Polymer until large strains, covering the elastoviscoplastic and hardening regimes. Well-established principles of parameter estimation in engineering science are used to found the identification procedure. It will be shown that three parameters only are necessary to model experimental tensile signals: the instantaneous ('Young's') modulus, the maximum relaxation time of a linear distribution (described with a universal shape) and a strain hardening modulus to describe the 'relaxed' state. The paper ends with an assessment of the methodology. Our results of instantaneous modulus measurements are compared with those obtained with other physical experiments operating at different temporal and length scales.

**Keywords:** inverse identification, instantaneous Young modulus, tensile behavior, semicrystalline polymer.

## 1. Introduction

During the last decades, one can observe a significant gap between recommended practices in the field of rheological parameter determination, relying on international standards, and the efforts made in the scientific community. On one hand, mechanical scientists develop more sophisticated models of behavior laws. On the other hand, experimentalists think about the appropriate metrology that can be associated to constitutive laws in order to precisely identify their parameters. The scientific field of inverse methods precisely deals with the mathematical basis of parameter estimation which can referred to as Model-Based Metrology (MBM). It has growth intensively since the eighties. MBM should nowadays be a very common framework when considering the metrology of physical parameters through pertinent physical modeling of experiments. It is necessary for a real enhancement of the quality in measured properties of materials which will, in turn, allows to investigate more deeply the link that can be made with the microstructural organization of materials. In order to illustrate this point of view, an example is given in this paper regarding the determination of the elastic modulus of polymers using the uniaxial tensile test. With respect to international standards, it rests upon the characterization of a linear elastic regime, i.e. proportionality between stresses and strains, at very small strains (short times). Among them, one can cite the ISO 527-1 and ASTM D 638 which suggest such a method leading to approximately the same results. Although this method is efficient and physically well-founded for metallic materials, it is not so satisfactory for polymers. Indeed, these materials quickly manifest a viscoelastic regime which takes place in the beginning of the tensile test. Therefore the obtained response does not show any true linear part. This clearly shows that an inverse approach of parameter estimation is required. This latter consists in the use of an optimization procedure based on uniaxial tensile test data and relying upon an adapted rheological constitutive model. The inverse problem is merely formulated as a least square optimization problem. It must be verified that the assessed parameters are not correlated and that their sensitivity is high enough in order to ensure a good quality of the identification. The aim of this paper is to show that this approach is not only possible but the only way to access pertinent measurement of the elastic modulus of materials like HDPE (High Density PolyEthylene) among SemiCrystalline Polymers (SCP). The sketch of the paper is as following: first, examples will be given of the observed mechanical behavior of HDPE in uniaxial tensile tests. True stress – true strain curves at different strain rates will illustrate the fact that it is very difficult and even false to measure correctly the elastic modulus of this polymer according to the recommended standards. In a second step, the rheological constitutive model will be presented in its most reduced version. This model will be shown able to describe different behavioral regimes (viscoelasticity, viscoplasticity, material hardening). In view of a parameter estimation problem, a reduced model means a model which meets the identifiability criteria i.e. a good adequation between the number of identifiable parameters and the relevance of the model with respect to the experiments (parsimony principle). In a third part, the identification procedure and the sensitivity analysis of the estimated parameters will be described. At last, the results of the identification will be discussed and compared to those given by the standards, and those measured by two other scientific techniques (ultrasonic device and nanoindentation tests).

## 2. <u>Tensile tests on HDPE : experimental results</u>

### 2.1 Material

The material tested in this work is HDPE (High Density Polyethylene, grade "500 Natural") produced by Röchling Engineering Plastics KG. Two different products (A and B) were manufactured under the same reference in an interval of six years and supplied in sheets (extrusion process). Information from the supplier indicates that the molecular weight and density are respectively of 500,000 g/mol and 0.935 g/cm$^3$. Differential scanning calorimetry gave a crystallinity index of 68 wt% for product A and 66 wt% for product B. Bone-shaped samples A were cut from a 6 mm thick sheet of polymer (along the extrusion direction " $A_{//}$ " (samples of reference in this paper) and perpendicularly to it " $A_{\perp}$ "). Some other samples were cut along the extrusion direction and machined to 3 mm thick (core samples " $A_c$ "). Samples B were cut along the extrusion direction only, from a 4 mm sheet produced six years earlier. Even if specimens A and B are supposed to be the same chemical material, this time interval between the dates of production has resulted in a different behavior. As the experiments performed on sample B at delivery and 6 years later lead to the same results, ageing can not be considered as responsible for this difference. Hence, the explanation probably lies in a different manufacturing process. The microstructure and the mechanical behavior of a polymer are clearly dependent on the cooling process which generates more or less crystalline phase and rules its distribution and organization in the bulk among the amorphous one. This unwanted difference will be seen to turn into real opportunity when considering the scope of the present analysis.

### 2.2 Video-controlled tensile test

All mechanical tests were performed on a servo-hydraulic MTS 810 load frame with Flextest SE electronic controller. A video-extensometer (VidéoTraction®) gives access to the elastoviscoplastic response of polymers under uniaxial tension. Local measurements of true strains are performed at the center of the neck. The measurement of the corresponding force enables us to construct the axial true stress – true strain curve of the sample (G'sell et al., 2002). Another important feature of the system is that it controls the servovalve of the machine in real time (through a feed-back loop) so that any desired input path for the true strain $\varepsilon_{11}$ can be imposed on the system. Seven dot markers are printed on the front face of the sample prior to deformation (Fig. 1), within a representative elementary frame of about 9 mm². The markers are black, nearly round and with a diameter of about 0.4 mm. Local deformations are measured between the dots according to Hencky's definition of the true strain:

$$\varepsilon = \ln\left(\frac{l}{l_0}\right) \tag{1}$$

The longitudinal deformations are interpolated through Lagrangian polynomials in order to get precisely the longitudinal strain value $\varepsilon_{11}$ in the section where the transversal strains are measured (Fig.1). Thanks to the special geometry of the specimen, these measurements are performed always where necking develops. A standard deviation of $10^{-4}$ is commonly achieved for the noise that corrupts this strain signal. Recent measurements using Digital Image Correlation (DIC system ARAMIS 3D from GOM Instruments) show that the longitudinal strain as measured by Videotraction lies within an error of less than 4% (accounting for both the 2D plane measurement bias, and the necking phenomenon which do not conserve the principal axis reference frame).The transverse true strain $\varepsilon_{22}$ is calculated by averaging the strains determined from the two pairs FC and CG (Fig. 1). The second transverse strain $\varepsilon_{33}$ in the third direction has been simply taken equal to $\varepsilon_{22}$ , with the assumption that the strain tensor complies with a transversally isotropic material (assumption checked by viewing both lateral faces of the specimen using our 3D DIC system). True stress takes into account the reduction of the cross-sectional area, $S < S_0$, undergone by the sample while it is stretched:

$$\sigma_{11} = \frac{F}{S} = \frac{F}{S_0} \exp(-\varepsilon_{22} - \varepsilon_{33}) = \frac{F}{S_0} \exp(-2\,\varepsilon_{22}) \tag{2}$$

A CCD camera mounted on a telescopic drive records images during the test and follows the elementary frame during deformation. The applied force $F$ is directly measured by a 5 kN load cell. An image analysis software computes the strains in real time and thus, the true stress can be determined according to eq. (2).

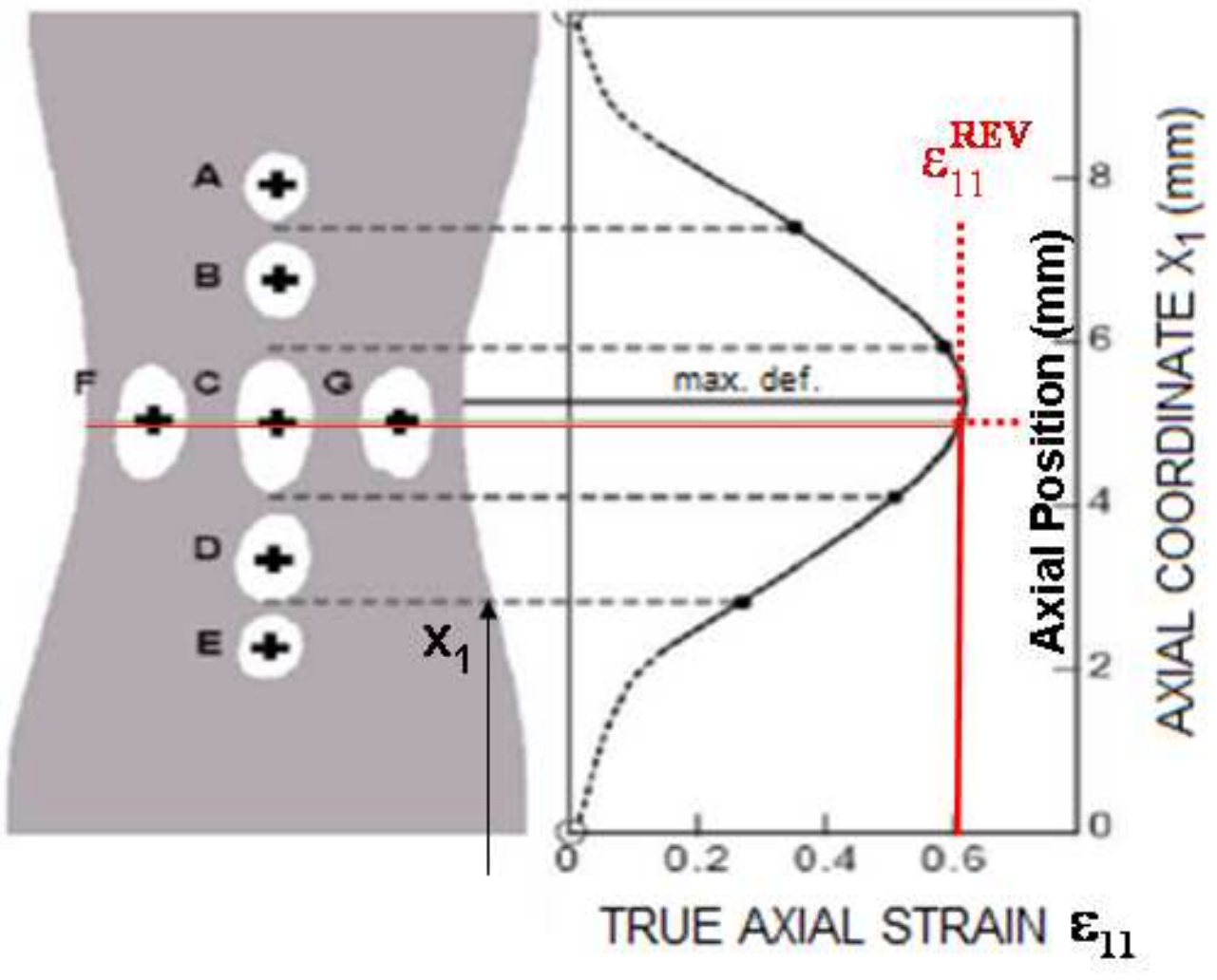

Figure 1: Principle of the true strain measurements as used by Videotraction system.

In the case where measurements of the transverse strains are not available, an isovolumic strain hypothesis can be used. This hypothesis is nearly met for our HDPE samples as proved with DIC measurements of the strain field. This approximation will be used later because it is an excellent way to demonstrate what an experimental bias consists in and how it can or cannot (in the present case) interfere with the identification of parameters. On Fig. 2, one can see that even if the transverse strain measurements exhibit some lack of reproducibility and of precision, the curves are nearly linear with approximately a -0.5 slope (Poisson coefficient $\nu$). Hence, we will consider the case where $\nu$ is set to 0.5.

This criterion is summarized as follows:

$$\nu = 0.5 \tag{3}$$

$$\varepsilon_{22}* = \varepsilon_{33}* = -\upsilon\ \varepsilon_{11} = -0.5\ \varepsilon_{11} \tag{4}$$

The stars upperscripts denote then the calculated transverse strains obtained from the unique measurement of $\varepsilon_{11}$ .

This leads obviously to a null volume strain:

$$\varepsilon_v = \varepsilon_{11} + \varepsilon_{22}* + \varepsilon_{33}* = 0 \tag{5}$$

and to a true stress:

$$\sigma_{11}* = \frac{F}{S_0}\exp\left(-\varepsilon_{22}* - \varepsilon_{33}*\right) = \frac{F}{S_0}\exp\left(\varepsilon_{11}\right) \tag{6}$$

Finally the observable $\sigma*$ is considered as depending only on the measurement of $\varepsilon_{11}$ . An example of representation of $\sigma$ and $\sigma*$ is given in Fig. 2. Fig. 3 represents true stress $\sigma*$ - true strain $\varepsilon_{11}$ curves obtained for three different strain rates $\dot{\varepsilon}_{11}$ .

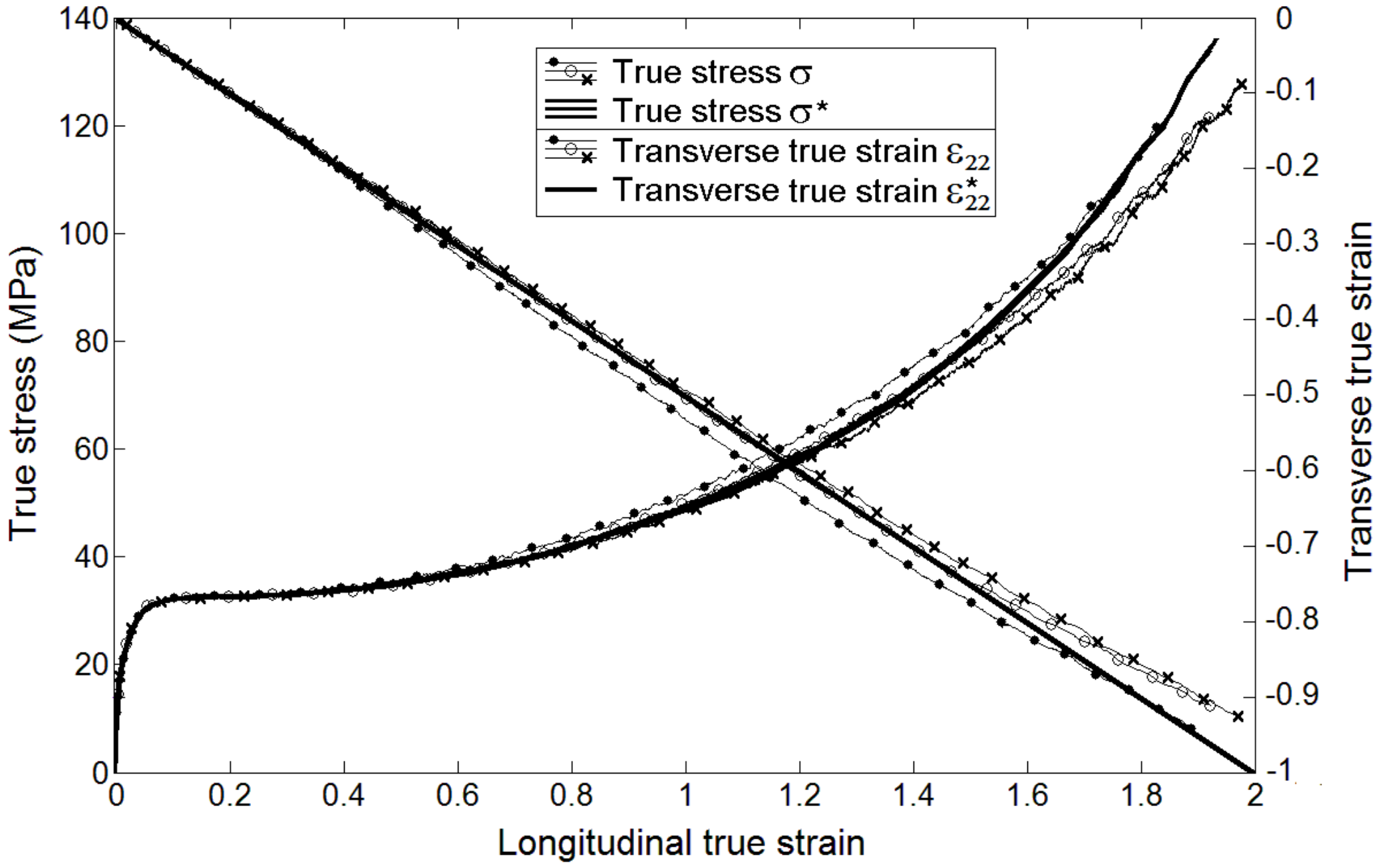

Figure 2: True stress $\sigma$ or $\sigma*$ (left axis) and transverse true strain $\varepsilon_{22}$ (right axis) versus longitudinal true strain $\varepsilon_{11}$ . Repeated experiments on specimen $A_{//}$ at $\dot{\varepsilon}_{11} = 0.005\ s^{-1}$ .

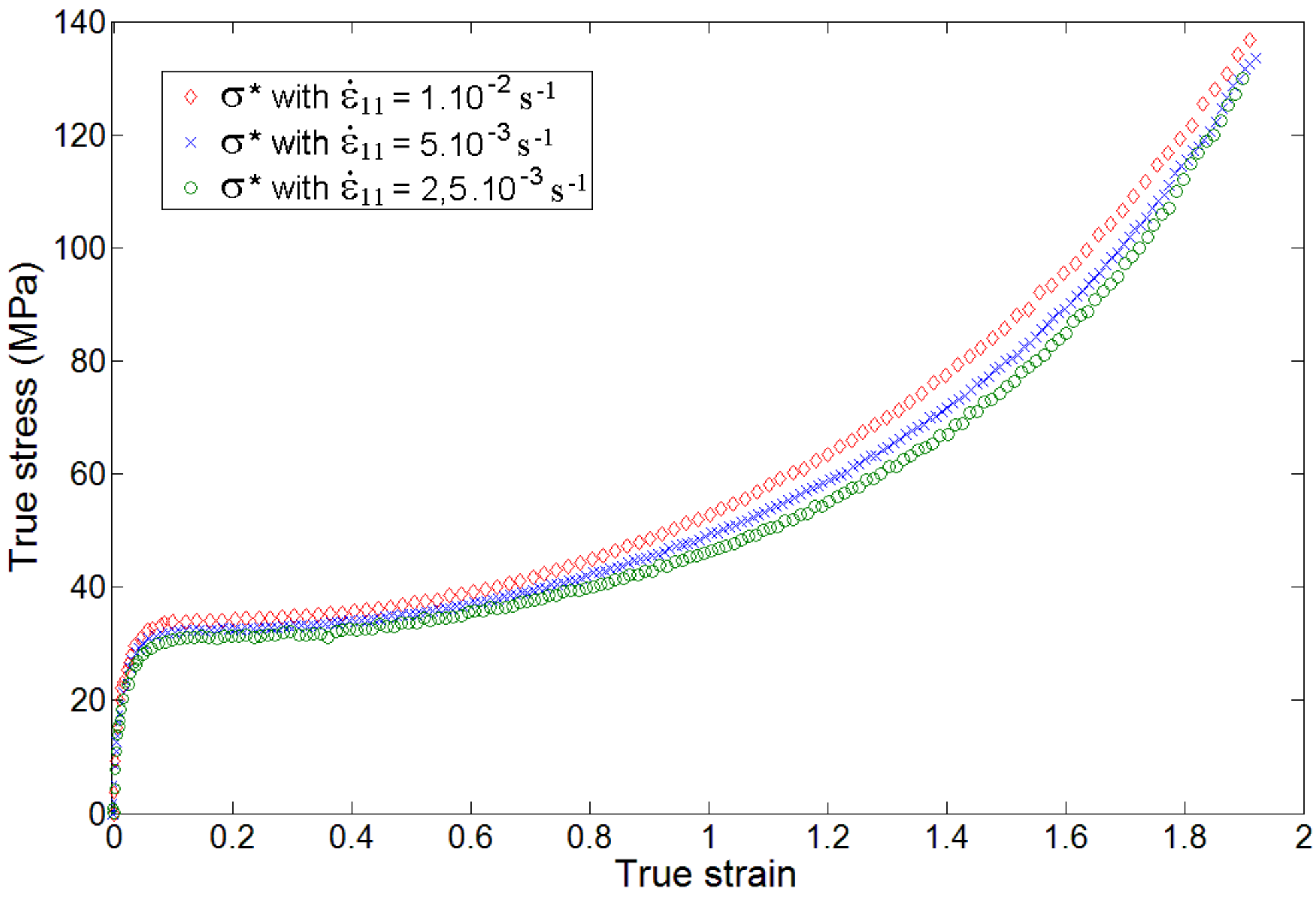

Figure 3: True stress $\sigma^*$ versus true strain $\varepsilon_{11}$ curves. (Specimen $A_{//}$ - Strain rates of $2.5\times10^{-3}s^{-1}$, $5\times10^{-3}s^{-1}$ and $10^{-2}s^{-1}$).

In addition, at large strains, an almost linear relation is observed (Fig. 4) when the true stress is plotted as a function of the strain deformation variable (Haward, 1993) defined as:

$$\varepsilon_{HT} = \lambda^2 - \lambda^{-1} = \exp(2\varepsilon) - \exp(-\varepsilon) \tag{7}$$

with λ, the extension ratio defined as:

$$\lambda = \exp(\varepsilon) = \frac{l}{l_0} \tag{8}$$

This strain variable naturally arises from a microscopic modeling of the entropic elasticity of a molecular network (Treloar, 1975). In this regime the hardening component of the stress behavior can be described with the following relation:

$$\sigma^{hard} = G\ \varepsilon_{HT} \tag{9}$$

where $G$ stands for the rubbery (or hardening or hyperelastic) modulus (in MPa).

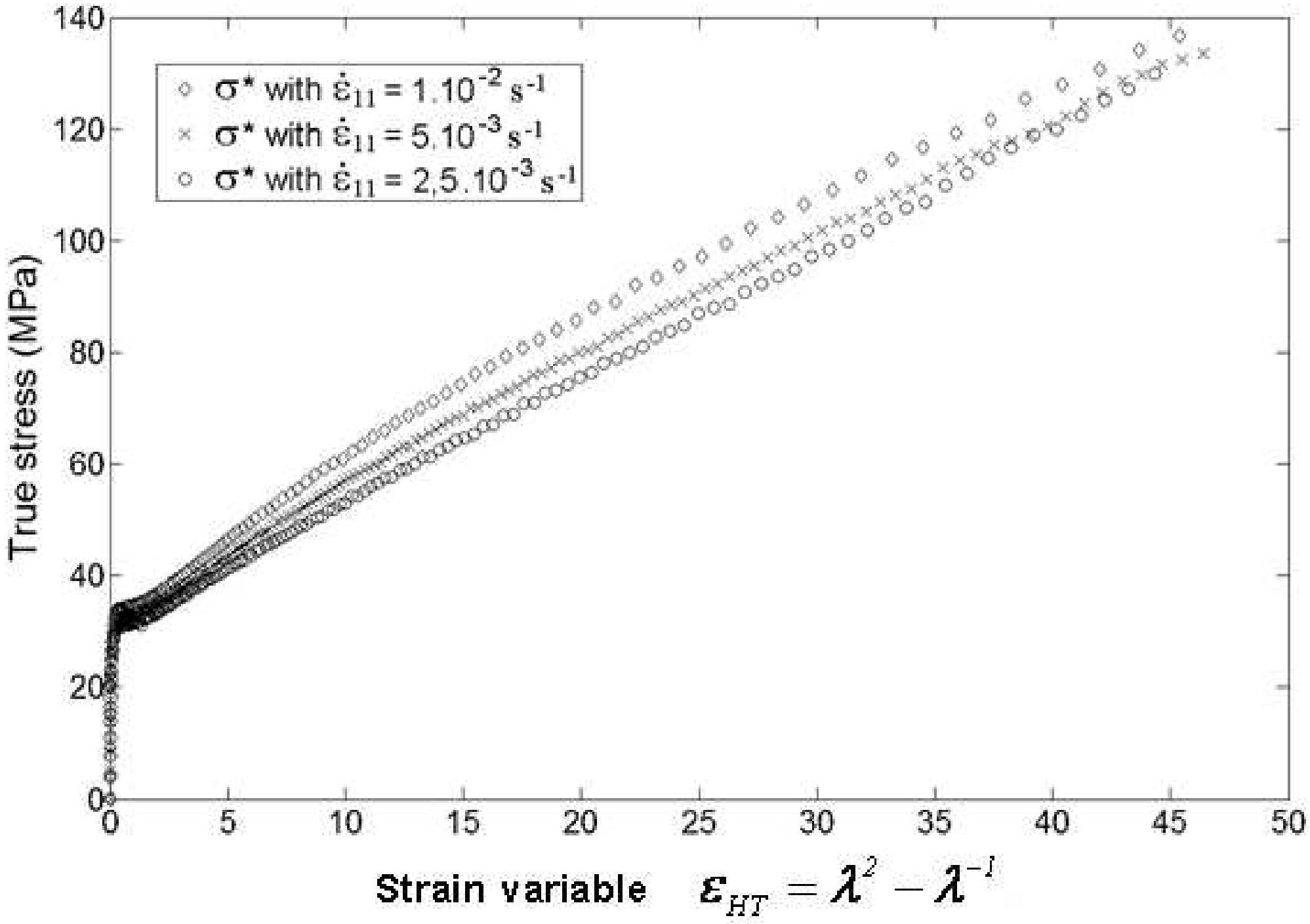

Figure 4: True stress $\sigma*$ versus strain variable $\varepsilon_{HT}$ (specimen $A_{//}$ - Strain rates of $2.5\times10^{-3}s^{-1}$, $5\times10^{-3}s^{-1}$ and $10^{-2}s^{-1}$).

### 2.3 Macroscopic behavior of HDPE in a tensile test

Various domains can be clearly identified on the true stress $\sigma*$- true strain $\varepsilon_{11}$ curves (Fig. 3). For very low strains, the mechanical behavior hardly exhibits linear elastic stage. A viscoelastic phase is observed until the apparition of a yield point for a strain value $\varepsilon_{11}$ close to 0.1 (referred to as $\varepsilon_{yield}$). At this point, the force reaches a maximum (first Considere condition) (Hiss et al., 1999) and is characterized by the development of necking. After the yield point, the mechanical response reveals the presence of a pseudo-plateau corresponding to the softening of the polymer followed by a plastic or flow regime. Referring to the recent paper by Haward (2007), the point where the differential nominal stress becomes positive (in terms of extension ratio), sometimes called the second Considere condition, is obtained for a value of 1.2. It corresponds to the stabilization necking condition and is considered as the transition between the viscoplastic flow and the strain-hardening regimes. The hardening phase is followed in the study up to strains of about $\varepsilon_{11} \approx 1.9$.

### 2.4 Measurements of the elastic modulus using standards

Standards ISO 527-1 and ASTM D 638 have been applied on experimental data (see Fig. 5 for an example of implementation of standard ISO 527-1) and the results for the elastic modulus are reported in table 1.

One can notice that different values are obtained if the nominal strain rate is changed (discrepancy of the order of 23 %). This should not be, because the elastic modulus is supposed to correspond to an instantaneous response with respect to the time constant of the excitation. This result is all the more unsettling for two reasons: firstly, because nowadays, parameter estimation in engineering science can not rely on such a poor procedure (adjusting a slope at the origin or between two values of strain). Secondly, because progress in material sciences will be possible only if physical parameters are identified according to a procedure that agrees with their definition. The elastic modulus is the instantaneous thermodynamic coefficient which links the stress and the strain in the corresponding state law and must be determined in full coherence with this status. The elastic modulus of polymers, as measured by standards, does not comply with an appropriate scientific definition and hence does not give satisfying results. We will demonstrate that using a well-thought strategy of identification of this intrinsic parameter, which relies on a correct constitutive model, can circumvent such problem.

| | ISO 527-1 | ASTM D 638 | Manufacturer |
|---|---|---|---|
| $\dot{\varepsilon}_N = 5$ mm/min | 1138 | 1141 | 1200 |
| $\dot{\varepsilon}_N = 6$ mm/min | 1475 | 1479 | |

Table 1. Elastic modulus of HDPE as measured by international standards for two different nominal strain rates (specimen $A_{//}$ - $MPa$ ).

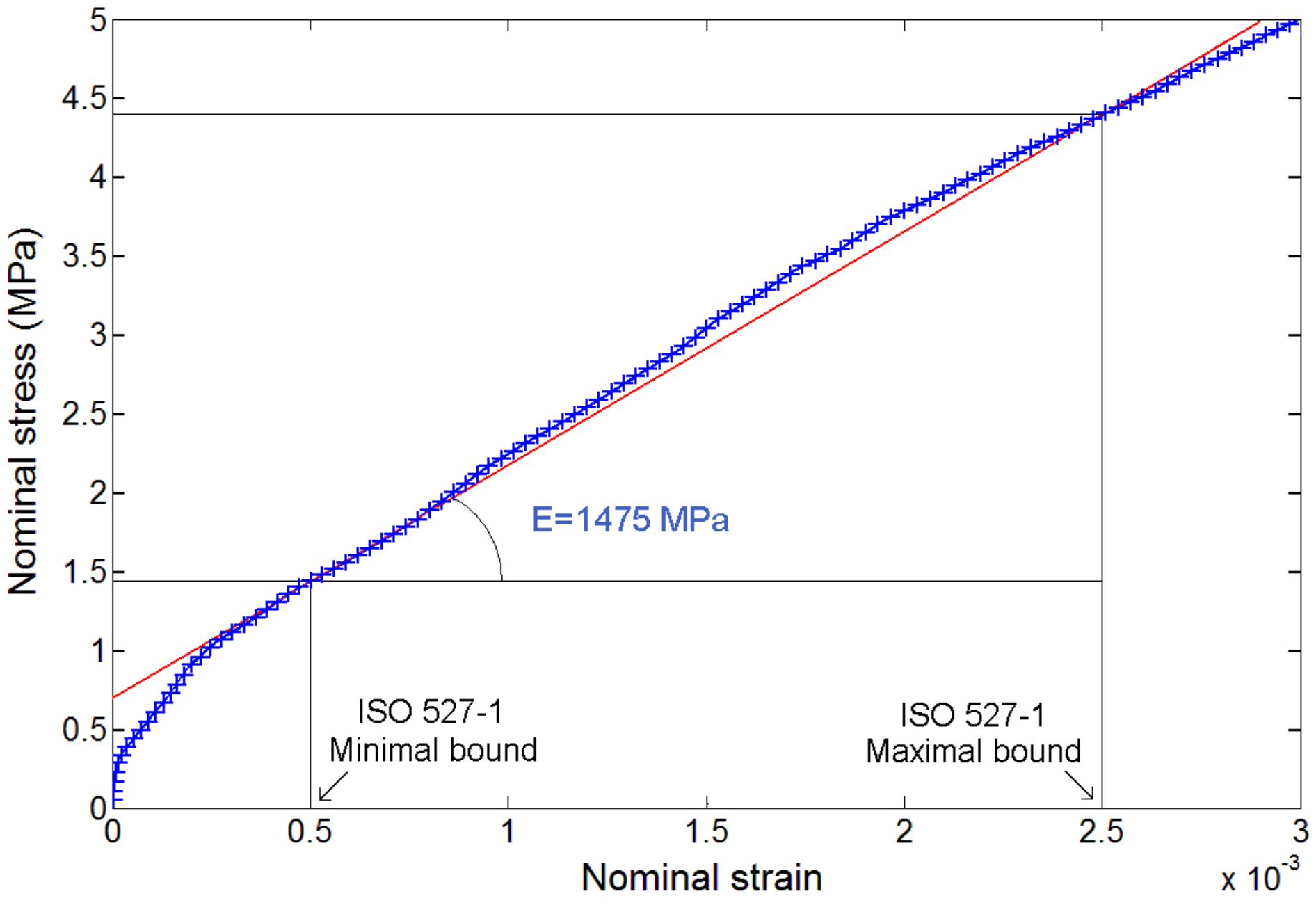

Figure 5: Nominal stress-nominal strain curve of HDPE at a strain rate of 6 mm/min (dotted curve) – The linear fit illustrates ISO 527-1 standard for determining elastic moduli.

## 3. Behavior model of HDPE and sensitivity analysis

### 3.1 Constitutive modeling

#### 3.1.1 General formulation

In order to characterize accurately and correctly the mechanical behavior of HDPE, a generalized viscoelastic constitutive model is necessary. The model used in this study is generally referred to as the DNLR approach (Distribution of Non-Linear Relaxations, see Cunat, 1991, 1996, 2001). It has been used to reproduce many different experiments (various polymers and loading conditions (Rahouadj et al., 2003; Mrabet et al. 2005)). This model relies on fundamental axioms of Thermodynamics of Irreversible Processes (T.I.P.) as stated for example by Callen (1985) or more recently by Kuiken (1994). The state laws of irreversible thermodynamics are classically decomposed into an instantaneous (or unrelaxed) component and a delayed one. In the following, the indexes $u$ and $d$ will refer respectively to the unrelaxed and the delayed components This decomposition is classical in other thermodynamical approaches that make use of internal variables (Maugin and Muschik 1994). In a tensile test where the strain is imposed, its dual thermodynamic variable is the stress and this variable constitutes the response of the system. The postponed response is described using a modal approach (lower indices $j$ in the equations denote the mode $j$) for the irreversible processes at stake in the microstructure. The existence of a modal basis to describe dissipative mechanisms has been proved in thermodynamics by Meixner (1949). It is simply described here by a first order kinetic model where the stress 'fluctuation' or difference between the modal true stress $\sigma_j$ and the modal relaxed stress $\sigma_j^r$ regresses according to time $\tau_j$. This time represents the relaxation time which governs the kinetics of each dissipation mode $j$. The overall true stress $\sigma$ is the sum of all the modal components $\sigma_j$ and the general relation of the DNLR formalism is written:

$$\dot{\sigma}=\dot{\sigma}^u+\dot{\sigma}^d=\sum_{j=1}^{N}\dot{\sigma}_j=\sum_{j=1}^{N}\left(E_j^u\,\dot{\varepsilon}-\frac{\sigma_j-\sigma_j^r}{\tau_j}\right) \tag{10}$$

$E_j^u$ corresponds to the modal unrelaxed modulus. $E_j^u$ and $\sigma_j^r$ can be simply defined by : $E_j^u = p_j^0 E^u$ and $\sigma_j^r = p_j^0 \sigma^r$ where $p_j^0$ is a coefficient that weights the modal component $j$ with respect to the global quantity. $\sigma^r$ refers to a relaxed state corresponding to the steady-state regime of the internal mechanisms, where the non equilibrium forces do not evolve in time. This thermodynamical state is clearly defined by $\dot{A}=0$ and named "iso-affin" state according to the recommendation of Prigogine (1946). $A$ is indeed the affinity state variable introduced by De Donder (1936) to take into account chemical reactions. It is used by mechanical scientists in solid rheology to describe internal reorganizations which take place in the matter under deformation processes (Kuiken, 1994). $E^u$ stands for the common elastic (Young's) modulus and hence, the modal weights must fulfill the normalized condition:

$$\sum_{j=1}^{N} p_j^0 = 1 \tag{11}$$

From eq. (10), we can easily infer that for a tensile test at very low strain rate, the true stress is expected to be very close to the relaxed stress (the non equilibrium forces are constant). According to this approach, the relaxed state can be seen as a pseudo-equilibrium which is still dynamic. It must not be mistaken for the equilibrium state (rigorously defined in T.I.P. by a zero-affinity state) which corresponds to the new microstructural conformation taken by the matter when the driving external forces are stopped. For example, in a same type of modeling (the V.B.O. approach due to Krempl, 2001), the distinction is clearly made although the same stress notation ($g$) and name ("equilibrium") is used for both: $g$ is considered as the equilibrium stress "when all time rates are zero" (must be equal to 0 in the absence of applied stress) and $g$ is considered as an equilibrium stress that has to evolve under application of an external force and is said to correspond to the stress "that must be overcome to generate inelastic deformation" (it corresponds to the variable $\sigma^r$ in the present model).

The DNLR approach in the form of eq. (10), offers then two entry points which can be adapted for different modelings: the spectrum of relaxation times and the relaxed state which will be discussed now.

### 3.1.2 Spectrum of relaxation times

Following the postulate of an equipartition of the created entropy accompanying regression of the fluctuations about an equilibrium state, Cunat (1991, 1996, 2001) proposed a universal spectrum that links each dissipative modal weight $p_j^0$ to its corresponding relaxation time $\tau_j$ (Eq. (12)).

$$p_j^0 = \frac{\sqrt{\tau_j}}{\sum_{j=1}^{N} \sqrt{\tau_j}} \quad (12)$$

The whole linear spectrum of relaxation times is generally distributed along a logarithmic scale in an interval [$\tau_{\min}$, $\tau_{\max}$]. Only one parameter is needed to describe this spectrum (in linear viscoelasticity). We generally retain the longest relaxation time $\tau_{\max}$ because it is more close to the physical perception of an experimenter. It is named $\tau_{\max}{}^T$ where the "*T*" upperscript denotes the Tensile phase of the test) The number of decades $d$ chosen for the temporal scale is a hyper-parameter which means that it is connected to the mathematical structure of the model: the results must be independent of the decade number once it is set to a given upper bound. A number of decades greater than or equal to 6 is generally required to reach convergence and to cover all possible relaxation times of the material (Eq. (13)).

$$\frac{\tau_{max}}{\tau_{min}} = 10^{d} \tag{13}$$

$N = 50$ dissipative modes are generally considered in order to build a quasi-continuous spectrum of relaxation times. To summarize, $d$ and $N$ will be considered as fixed (model structure parameters having a null sensitivity on the model once they are set to appropriate physically sounded values) and the linear spectrum of relaxation times $\tau_j$ is defined once parameter $\tau_{max}{}^{T}$ is known, according to the following equation (Eq. (14)):

$$\tau_j = \tau_{max}\ 10^{-\left(\frac{N-j}{N-1}\right)d} \tag{14}$$

To account for a non-linear spectrum, a multiplying factor (shift-factor) is classically used and denoted by $a(T, t, ...)$. It shifts the time scale according to some dependent law of other variables like temperature $T$ (Arrhenius or V.F.T. law), deformation rate $\dot{\varepsilon}$ (power-law form), stress tensor $\underline{\underline{\sigma}}$ (to recover plastic domains)... In this form, the DNLR model rigorously corresponds to the Biot model for viscoelasticity (Biot, 1958) or to a generalized Maxwell model (Tschoegl, 1989), except that a recursive-type relation links both the relaxation times and weights (André et al., 2003).

### 3.1.3 Relaxed state model adopted for HDPE under tensile tests

The second entry point for the modeling lies in the description chosen for the relaxed stress $\sigma^{r}$. The most basic form consists in assuming the linear form $\sigma^{r} = E^{r}\,\varepsilon$ which can produce an elasto-visco-plastic behavior but with only linear hardening in plastic regime. As we saw in section 2, a constitutive law for HDPE must comply with a more complex continuous modeling from the viscoelastic to viscoplastic and hardening behaviors. Hence, a different modeling must be considered for the relaxed state. A SCP in highly deformed state is generally considered as a tridimensional network of rubber type phase with entanglements. A first model for elastomers hardening considered by Wang and Guth (1952) and then modified by Arruda and Boyce (1993) put forward cells with eight sub-chains by node which is close to a random distribution of the chains in a real material and does not favour any spatial direction. By considering the works of Haward & Thackray (Haward, 1993) and Treolar (1975), Arruda and Boyce have shown that during hardening, the "back stress" (or anisotropic resistance to chain alignment) can be described by the following formula:

$$\sigma^{\text{backstress}} = \frac{N_0 k_B T}{3\lambda_c}\sqrt{n}\; L^{-1}\left(\frac{\lambda_c}{\sqrt{n}}\right)\varepsilon_{HT} \tag{15}$$

with $\lambda_c = \sqrt{\left(\exp(2\,\varepsilon) + 2\exp(\varepsilon)\right)/3}$ and where, $N_0$ stands for the density of chains per unit volume, $n$ is the number of segments per chain, which controls the behavior at large strains until the extreme extensibility of the network, $k_B$ is the Boltzmann constant and $T$, the temperature. $L^{-1}$ represents the inverse Langevin function.

For a given temperature, eq. (15) requires the knowledge of the two parameters $N_0$ and $n$. It can be reduced to the simplified version:

$$\sigma^{\text{backstress}} \approx G\;\varepsilon_{HT} \tag{16}$$

where $G$ is the single parameter, referred to as the hardening modulus.

According to Krempl (2001), the term "back stress" is often used in the literature but is not relevant for describing the subtle differences existing between equilibrium stress and overstress (relaxed state in our approach) and already pointed out in section 3.1.1. Besides, Negahban (2006) considers the "back stress" as the delayed response (or inelastic response) and that it can correspond to the “back stress” involved in the Arruda & Boyce models (Arruda and Boyce, 1993). In the framework of Irreversible Thermodynamics based on the concept of Affinity, such considerations are aimless. It makes clear the distinction between actual (or unrelaxed), relaxed, and equilibrium states. Turning now our attention to the experimental results presented in section 2 (Fig. 4), it is clear that the true stress plotted as function of the Haward-Thackray strain variable $\varepsilon_{HT}$ exhibits an almost linear relation with constant slope, whatsoever the strain rate. This fact has been already highlighted in other works (Van Melick et al., 2003). Therefore in this paper, we consider that the “back stress” in Arruda & Boyce model could be associated to the relaxed state of our model and thus, eq. (19) will be considered. Indeed, the Haward-Thackray variable puts forward the behavior at large strains, when all the relaxation times have been depleted.

$$\sigma^{r} = G\ \varepsilon_{HT} \tag{17}$$

Finally the reduced model that will be considered in this study, contains only three parameters: $E^u$, $G$ and $\tau_{\max}{}^{T}$ and is described by eq. (20).

$$\dot{\sigma} = \sum_{j=1}^{N} \dot{\sigma}_j = \sum_{j=1}^{N} \left( E_j^{u}\ \dot{\varepsilon} - \frac{\sigma_j - G\ \varepsilon_{HT}}{\tau_{\max}\ 10^{-\left(\frac{N-j}{N-1}\right) d}} \right) \tag{18}$$

The relevance of this model will be proven now through the sensitivity analysis and parameter identification procedure (section 4).

### 3.2 Sensitivity analysis fundamentals

Let us consider $y^m(t)$ as the measured output (the stress) of a system (our material specimen), and $y(\beta,t)$ the theoretical stress (output) of the behavioral model with parameter vector $\beta$ of dimension $p$, representing in our case the $p$ constitutive material parameters. We can then define the output error $e(t)$ (or the residuals) by the following formula:

$$e(t) = y^m(t) - y(\beta,t) \tag{19}$$

The estimator aims at minimizing this output error. A least square criterion can be used classically which is written as follows:

$$E_{LS} = \sum_{i=1}^{q} \left( y^m(t_i) - y(\beta,t_i) \right)^2 \tag{20}$$

where the summation is made over the $i^{th}$ data points corresponding to the successive acquisition times $t_i$ ($q$ stands for the total number of experimental data points).

The minimization of the criterion is done when its derivative with respect to parameters $\beta_j$ ($j = 1...p$) is null:

$$\forall j \in [1, p], \ \frac{\partial E_{LS}}{\partial \beta_j} = 0 \quad \rightarrow \sum_{i=1}^{q} \left( \frac{\partial y(\beta,t_i)}{\partial \beta_j} \left[ y^m(t_i) - y(\beta,t_i) \right] \right) = 0$$
$$\rightarrow \sum_{i=1}^{q} \left( X_j(\beta,t_i) \left[ y^m(t_i) - y(\beta,t_i) \right] \right) = 0 \tag{21}$$

From this equation, the sensitivity coefficient vector $X_j$ associated to parameter $\beta_j$ can be clearly recognized as:

$$X_j\left(\beta,t_i\right)=\frac{\partial y(\beta,t_i)}{\partial\beta_j} \tag{22}$$

Sensitivity coefficients express how much a model reacts to some small variation of the parameters. Sensitivity coefficients have a fundamental role in the conditioning of the inverse parameter estimation process and, as a consequence, upon the errors made in the estimations (confidence bounds). It is obvious that large sensitivities are sought for when designing an experiment in view of metrological purposes. Note that in almost all cases, sensitivity coefficients are non linear functions of the parameters themselves (when the model $y$ is a non linear function of the $\beta_j$).

Switching to a matrix formulation and defining vectors $Y^m$ and $Y$ such as:

$$Y^m=\begin{bmatrix} y^m\left(t_1\right)\\ y^m\left(t_2\right)\\ \cdots\\ y^m\left(t_q\right)\end{bmatrix} \text{ and } Y=\begin{bmatrix} y\left(\beta,t_1\right)\\ y\left(\beta,t_2\right)\\ \cdots\\ y\left(\beta,t_q\right)\end{bmatrix} \tag{23}$$

makes the minimization process now expressed as

$${}^tX\left(Y^m-Y\right)=0 \tag{24}$$

with $X$ , the (rectangular) sensitivity matrix whose rows are for each observation time $t_i$ and columns are for each parameter $\beta_j$:

In the case of a linear model with respect to the parameters (linear estimation problem), we have:

$$Y=X\beta \tag{25}$$

where the matrix of the sensitivity coefficients does not depend on the parameters.

The estimated parameter vector, denoted by $\hat{\beta}$, corresponds to the value reached by $\beta$ when the criterion is minimized. Therefore by using (25) we can then rewrite the equation (24) into:

$$ {}^{t}X\left(Y^{m}-X\hat{\beta}\right)=0 \tag{26}$$

Relation (26) can be inverted in order to obtain the expression of $\hat{\beta}$ in the case of a linear estimation problem only:

$$\hat{\beta}=\left({}^{t}X\ X\right)^{-1}\ {}^{t}X\ Y^{m} \tag{27}$$

Of course, most parameter estimation problems are not linear and require an iterative linearizing procedure. It is obtained by developing the solution (at rank *n+1*) in the neighborhood of the solution obtained for the prior iteration (rank *n*):

$$Y^{(n+1)}=Y^{(n)}+X^{(n)}\ \left(\hat{\beta}^{(n+1)}-\hat{\beta}^{(n)}\right) \tag{28}$$

By combining relation (24) written at rank *n+1* for parameters estimated at rank *n*: ${}^{t}X^{(n)}\left(Y^{m}-Y^{(n+1)}\right)=0$ and relation (28), we obtain the following relation of recurrence between estimated parameters at rank *n+1* and rank *n*:

$$\hat{\beta}^{(n+1)}=\hat{\beta}^{(n)}+\left({}^{t}X^{(n)}X^{(n)}\right)^{-1}\ {}^{t}X^{(n)}\ \left(Y^{m}-Y^{(n)}\right) \tag{29}$$

which defines the iterative procedure that can be used for estimating the parameters (Gauss-Newton algorithm).

In the followings, we discuss the statistical properties of the estimator, which depends on the noise $\varepsilon(t)$ of the signal. If the theoretical model is assumed unbiased (perfect) then we have:

$$Y^m(t) = Y(t,\beta) + \varepsilon(t) \tag{30}$$

If classical statistical assumptions are made regarding the experimental noise $\varepsilon(t)$ on the measured signal (the stress) (Beck et al., 1977), it is possible to get an estimation of the errors that can be made in the estimation process for the different parameters. These hypotheses are:

- zero mean value of the signal in the absence of excitation, which corresponds to a zero expectancy for the noise (expected value $E(\varepsilon) = 0$) ;
- constant variance or standard deviation of the noise: $V(\varepsilon) = \sigma_0^2$ .

In the case of a 1st order linearized estimation, its expected value can be proved to be:

$$E(\hat{\beta}) = \beta \tag{31}$$

This means that there is no error or bias made on the identified parameters.

The variance-covariance matrix Δ on the estimated parameters (generalization of the scalar-valued variance to higher dimension) naturally involves the sensitivity coefficients. It is calculated as $\Delta = E\left[\left(\hat{\beta} - E\left(\hat{\beta}\right)\right)\ ^t\left(\hat{\beta} - E\left(\hat{\beta}\right)\right)\right] = \sigma_0^2\left({}^tX\ X\right)^{-1}$ which in expanded form gives:

$$\Delta_{rs} = \begin{bmatrix} V(\beta_1) & \text{cov}(\beta_1,\beta_2) & \ldots & \text{cov}(\beta_1,\beta_p) \\ \text{cov}(\beta_1,\beta_2) & V(\beta_2) & \ldots & \text{cov}(\beta_2,\beta_p) \\ \ldots & \ldots & \ldots & \ldots \\ \text{cov}(\beta_1,\beta_p) & \text{cov}(\beta_2,\beta_p) & \ldots & V(\beta_p) \end{bmatrix} \tag{32}$$

A stochastic analysis has been made which consists in calculating theoretically (according to a given noise and given set of parameters) the variance-covariance matrix through eq. (32). This matrix is symmetric squared with dimension equals to the number of parameters. The diagonal terms correspond directly to the variance of each parameter $V(\beta_j)$. They can be used to determine the error made on each parameter. We will present this error (expressed in %) as:

$$Err(\beta_j) = \frac{\sqrt{V(\beta_j)}}{\beta_j} \tag{33}$$

The off-diagonal terms can be used to calculate the correlation coefficients $\rho_{mn}$ which express the degree of correlation of the parameters :

$$\rho_{rs} = \frac{\mathrm{cov}(\beta_r, \beta_s)}{\sqrt{V(\beta_r)}\sqrt{V(\beta_s)}} \tag{34}$$

The values for $|\rho_{rs}|$ lie between 0 and 1. In the case of a model with strongly correlated parameters, the correlation coefficients are 'close' to 1 which means that two columns of the sensitivity matrix $X$ are nearly proportional. The resulting confidence bounds interval for two correlated parameters are therefore generally very high. This means that a large number of solutions exist for these two parameters to allow for a good fit of the experimental curve. A deterministic algorithm used for the minimization process (like the steepest gradient technique) is consequently very sensitive to the initial guess made for the parameters. A strategy to produce first approximate estimates using physical background is highly recommended. But still, the estimation problem is ill-posed and indicates to the experimentalist that the involved physical description is probably not appropriate and must be changed.

In the followings, the identifiability of the model parameters will be analyzed through array $\tilde{\Delta}$ which combines the variance-covariance matrix and the correlation matrix. $\tilde{\Delta}$ puts the main diagonal of $\Delta$ on its main diagonal and the correlation coefficients on the off-diagonal terms.

$$\tilde{\Delta}_{rs} = \begin{bmatrix} Err(\beta_1) & \rho_{12} & \dots & \rho_{1p} \\ \rho_{12} & Err(\beta_2) & \dots & \rho_{2p} \\ \dots & \dots & \dots & \dots \\ \rho_{1p} & \rho_{2p} & \dots & Err(\beta_p) \end{bmatrix} \tag{35}$$

When there is no model bias (perfect agreement between the conditions of the experiment and the model based on it), the relation between the estimated parameter vector for the non linear estimation problem and its ‘exact’ value can be given at convergence with the following formula:

$$\hat{\beta} = \beta + \left( {}^{t}X \ X \right)^{-1} \ {}^{t}X \ \varepsilon \tag{36}$$

In that case, the residuals $e(t)$ (the difference between the model and the data) correspond exactly to the noise. Their standard deviation corresponds then to the experimental standard deviation of the noise and the residuals remain unsigned (no large fluctuation around the zero level)

The matrix $\tilde{\Delta}$ is a good tool to investigate identifiability conditions of a parameter estimation problem. Besides, use can be made of the normalized sensitivity coefficients:

$$\tilde{X}_j = \beta_j \frac{\partial y(\beta_j, t)}{\partial \beta_j} \tag{37}$$

in order to check graphically the level of sensitivity and possible correlations between parameters (different intervals of the independent variable, the time $t$ here may be advantageously considered). These coefficients will be dimensionally homogeneous to the signal itself, the stress, and will be specified in MPa.

## 4. Identification procedure

### 4.1 Adjustment results

The model $\sigma^*(\varepsilon)$ described previously is now applied to experimental data in order to identify the following parameter vector: $\beta = [E^u, \mathrm{G}, \tau_{\max}{}^T]$. Computations are made with Matlab Software and use is made of the Levenberg-Marquardt and/or Simplex algorithms to minimize the least square criterion. Once the convergence is reached and the parameters determined, the sensitivity coefficients at this point of the parameters space will be calculated along with the resulting optimal confidence bounds. The initial values used as starting guesses for the parameter vector are determined according to physical basis (a general rule that should be followed : if the literature uses the wording "knowledge" model in the field of system identification, this means that the researcher should have ideas or at least basic methods to roughly determine its parameters). An initial instantaneous modulus can be derived from the slope at the origin of the stress-strain curve (or using the standards). An initial value for the maximum relaxation time can be simply derived by assuming a pure exponential-type behavior of the rising part of the curve as a result of a step input. For the experiment performed at a strain rate of $5\times10^{-3}s^{-1}$ for example, a decaying strain of the order of 0.04 can be deduced from the rule of tangents (or logarithmic plot) and leads to a typical time of about 8 seconds. Finally, a "visual" linear regression applied on the experimental data of Figure 3 gives a first rough estimate of the order of magnitude of $G = 110/50 = 2.2$ which later will be seen also as a perfect initialization value of the minimization algorithm.

Two different identification intervals can be considered. In case where small strains levels ($\varepsilon \le \varepsilon_{yield} \approx 0.13$ – Interval I) are considered, the way the relaxed state is modeled has no influence on the solution. Only parameters $E^u$ and $\tau_{\max}{}^T$ can be identified. This means that the modeling can not be sensitive to the hardening phase as the polymer is still in the viscoelastic regime. Identifications performed either on $\sigma$ or $\sigma^*$ lead to the same results. On Fig. 6, the experimental data are plotted along with the model for the optimal identified values (insert of the figure). Leaving aside the first points of the curve, highly dependent on the feedback loop control efficiency, the identification residuals lie in the [-1, 1] MPa range, which represents a maximal discrepancy of 3% with respect to the yield stress value (of the order of 33 MPa). If the identification interval is extended to very large strains ($0<\varepsilon<2$ - Interval II), then the complete model as to be used to account for the hardening.

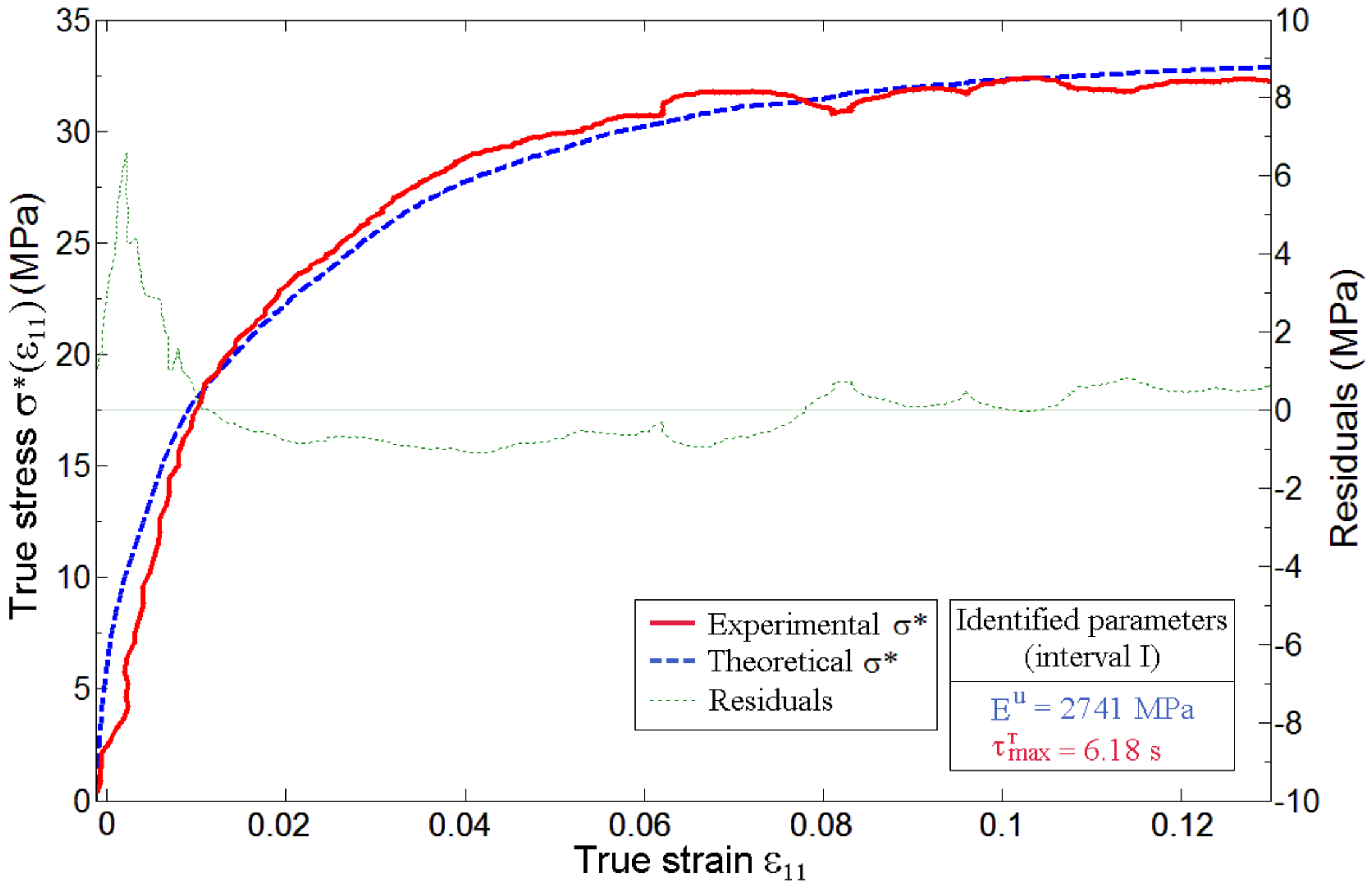

Figure 6: Experimental data, fitted curve and residuals obtained when identifying parameters $E^u$ and $\tau_{\max}{}^T$ on interval I (Specimen $A_{//}$ - Strain rate of $5\times10^{-3}s^{-1}$).

Figs. 7, 8 and 9 give the experimental and fitted curves obtained for specimen $A_{//}$ for three different strain rates. The good agreement between the model and the experimental tensile curves is obvious. The residuals, that is the difference between experimental data and recalculated model for the optimally identified parameters, are also plotted (with magnification) on the same figures. In the ideal case where the model would perfectly represent the true behavior of the material and where the experimental set-up would be in perfect agreement with the assumptions of the model (e.g., perfect ramp excitation, exact sensors …), these residuals would be unsigned and distributed around the zero-value. They should in fact represent the only measurement noise made on $\sigma^*$ and $\varepsilon_{11}$. The experiment carried out at the lowest strain rate (Fig. 7) is the closest to this picture. In that case, the parameters are perfectly identified and the confidence interval is mainly due to the contribution of the noise (eq. (35)). But still, the residuals show a signed character, which means that some bias exist (inadequation between model and data) that induces an additional error on the estimated parameters (biased estimations). The existence of a bias generally helps the experimentalist to improve the couple model-experiment. If the bias remains limited (of the order of the noise level) then the parameters can be considered as well-identified. Well-identified means here that according to the sensitivity analysis made in section 3.2, the estimation process produces the unique set of parameters (the optimal one) minimizing the residuals in the least-square sense. This means that in case of no bias (in the data, in the model) but in the presence of the same amount of noise, the variances on the un-biased parameter estimates will be lower-bounded by the estimated variances calculated in a stochastical manner (eq. 37). Here, the bias seems more important when the strain rate increases. For a strain rate of $10^{-2}s^{-1}$, the residuals lie in the [-10%, 10%] range of the current stress values. A possible origin of this bias may come, in such mechanical tensile tests, from a non ideal input command. In our case, the input ramp $\varepsilon(t) = \dot{\varepsilon}\, t$ is very good, due to the special care taken in the selection of the PID parameters used by the control feed-back loop. Furthermore, the real input signal is used in the calculation of the model response. Here, the bias is more likely due to an approximation introduced in the model for the material behavior. This approximation comes from the use of $\sigma^*$ instead of $\sigma$ (see eqs. (2) and (6)) which avoids introducing the measurements made on the transversal strain to calculate the cross-section variations of the specimen with respect to time. The bias is also due to the approximation used in the relaxed state modeling (Haward-Thackray relation, eq. (17)) as evidenced by the curves shown on fig. 4 (the linear behavior is not perfect). Trying to reduce the bias by tracking defects with respect to the idealized experimental conditions is the first thing to do. Then a refinement of the model can be legitimate as a second step. Otherwise, a refinement of the model stimulated by biased data can be dramatic in view of the Parameter Estimation Problem. For example, trying to use here the Arruda & Boyce model of eq. (15) will introduce additional unnecessary parameters which may in return distort the parameter estimation problem (poor identifiability conditions). Here, as far as the residuals remain small with respect to the signal to noise ratio and because the aim of the paper is mainly devoted to show how this model can lead to better estimations of the instantaneous Young modulus, these attempts are not presented here.

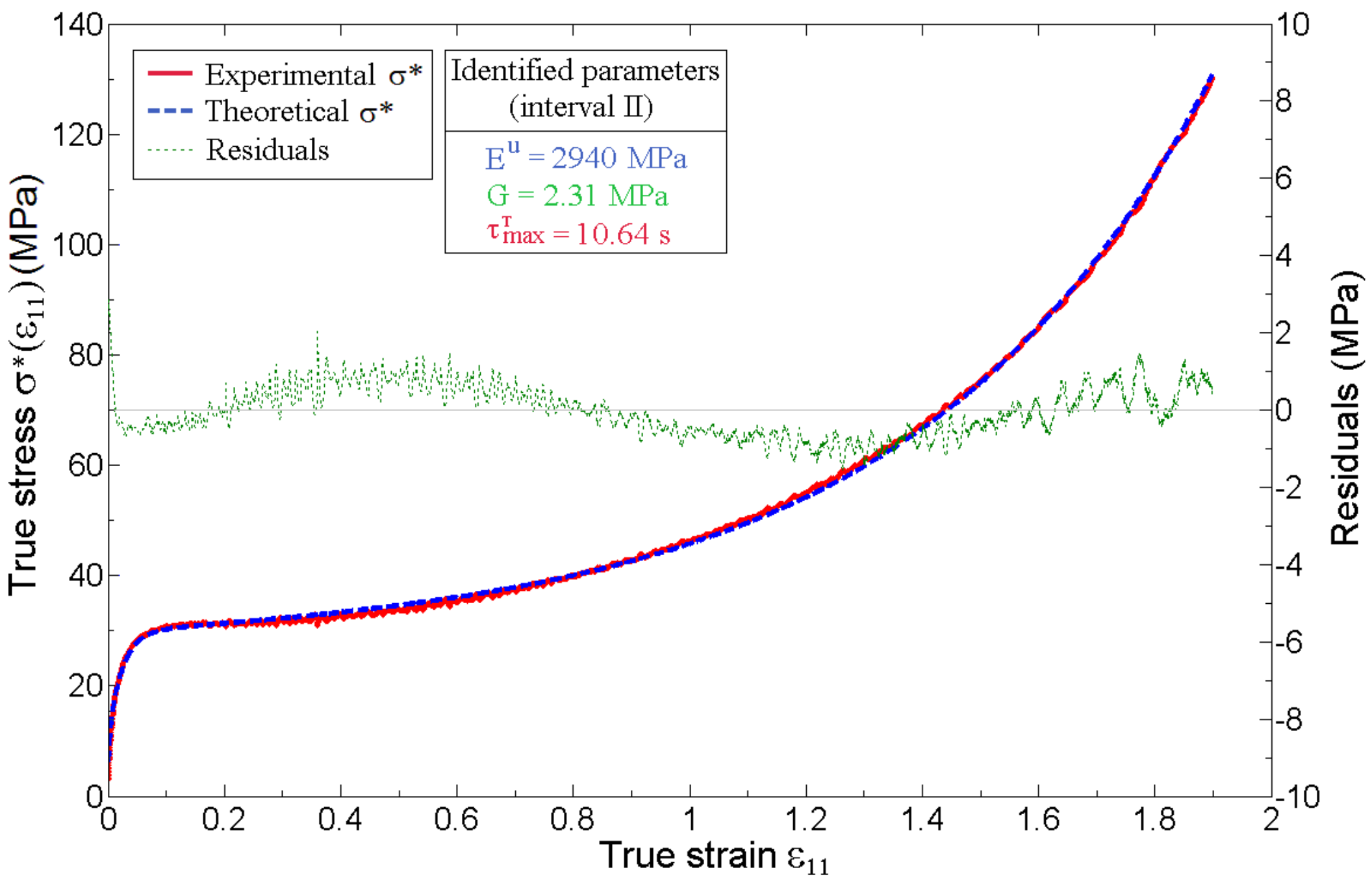

Figure 7: Experimental data, fitted curve and residuals obtained when identifying parameters $E^u$, G and $\tau_{\max}{}^T$ on interval II (Specimen $A_{//}$ - Strain rate of $2.5\times10^{-3}s^{-1}$).

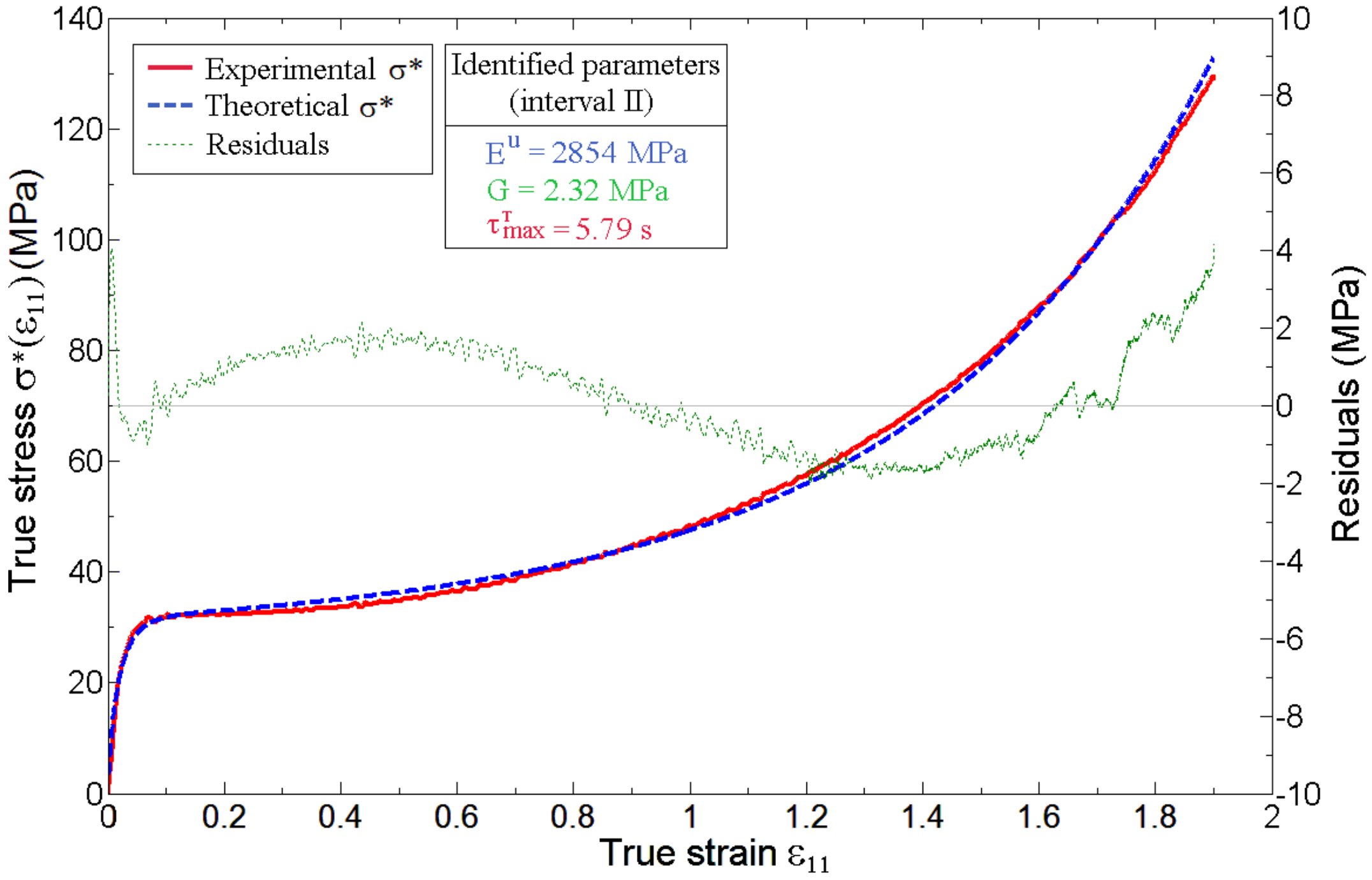

Figure 8: Experimental data, fitted curve and residuals obtained when identifying parameters $E^u$, G and $\tau_{\max}{}^T$ on interval II (Specimen $A_{//}$ - Strain rate of $5\times10^{-3}s^{-1}$).

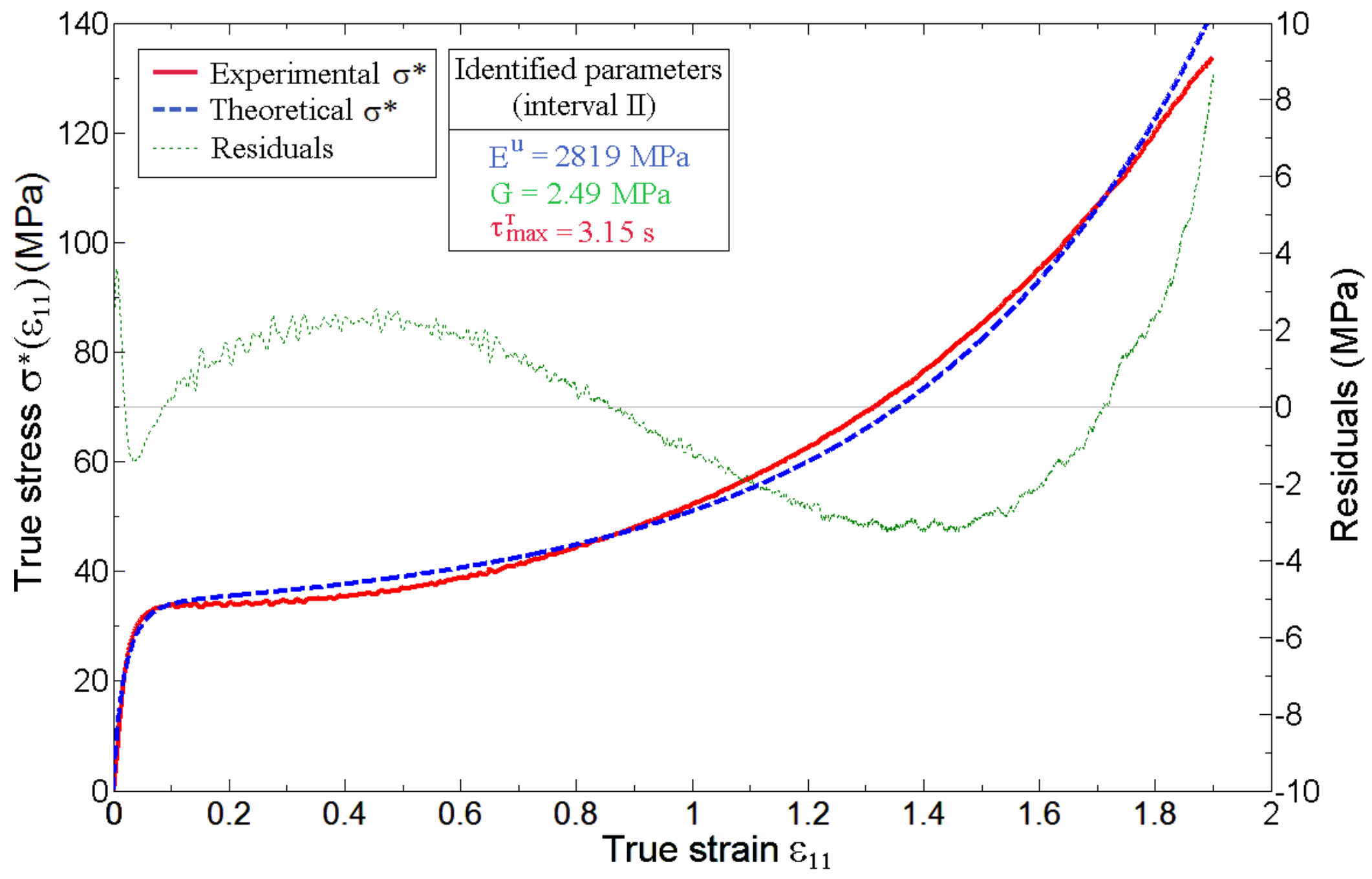

Figure 9: Experimental data, fitted curve and residuals obtained when identifying parameters $E^u$, G and $\tau_{\max}{}^T$ on interval II (Specimen $A_{//}$ - Strain rate of $10^{-2} s^{-1}$).

It is interesting also to discuss the results of the identification procedure when the tensile test is followed by a relaxation (initiated around a strain of 1.9). The reduced model can be used here too but with some modifications. Indeed, when relaxation starts, the material is highly deformed. This new excitation applies on a material which is different and implies to "reset" the knowledge we have on the microstructure. The relaxed state and the times spectrum must be described differently. For the relaxed state a simple constant stress value is assumed to be reached at long times and noted $\sigma^r = \sigma_\infty$ ("equilibrium" state at rest). The spectrum conserves its properties (linearity, number of decades and number of modes) but it is simply shifted in time according to a new maximum relaxation time noted $\tau_{\max}{}^R$, which can be considered as a new characteristic of the material (in fibrillar state for such a high deformation when relaxation is initiated). Note that in view of our constitutive model, the instantaneous (or unrelaxed) modulus has not to be considered different. For the tensile-relaxation test, the parameter vector that is considered is now $\beta = [E^u, \mathrm{G}, \tau_{\max}{}^T, \sigma_\infty, \tau_{\max}{}^R]$. The results of the identification with this modified (but still reduced) model are shown on Fig. 10 for specimen B. It can be seen that the agreement is very good both in the tensile and relaxation stages, with the same $E^u$ value. Physically, this means that elastic (instantaneous) properties of the SCP are the same in the initial bi-phasic microstructure and in the fibrillar state (when unloading is triggered). It is emphasized again that this reduced approach is made for a pure metrological purpose. This model would have to be made more complex to account for more sophisticated loading paths such as multiple loading-unloading stages.

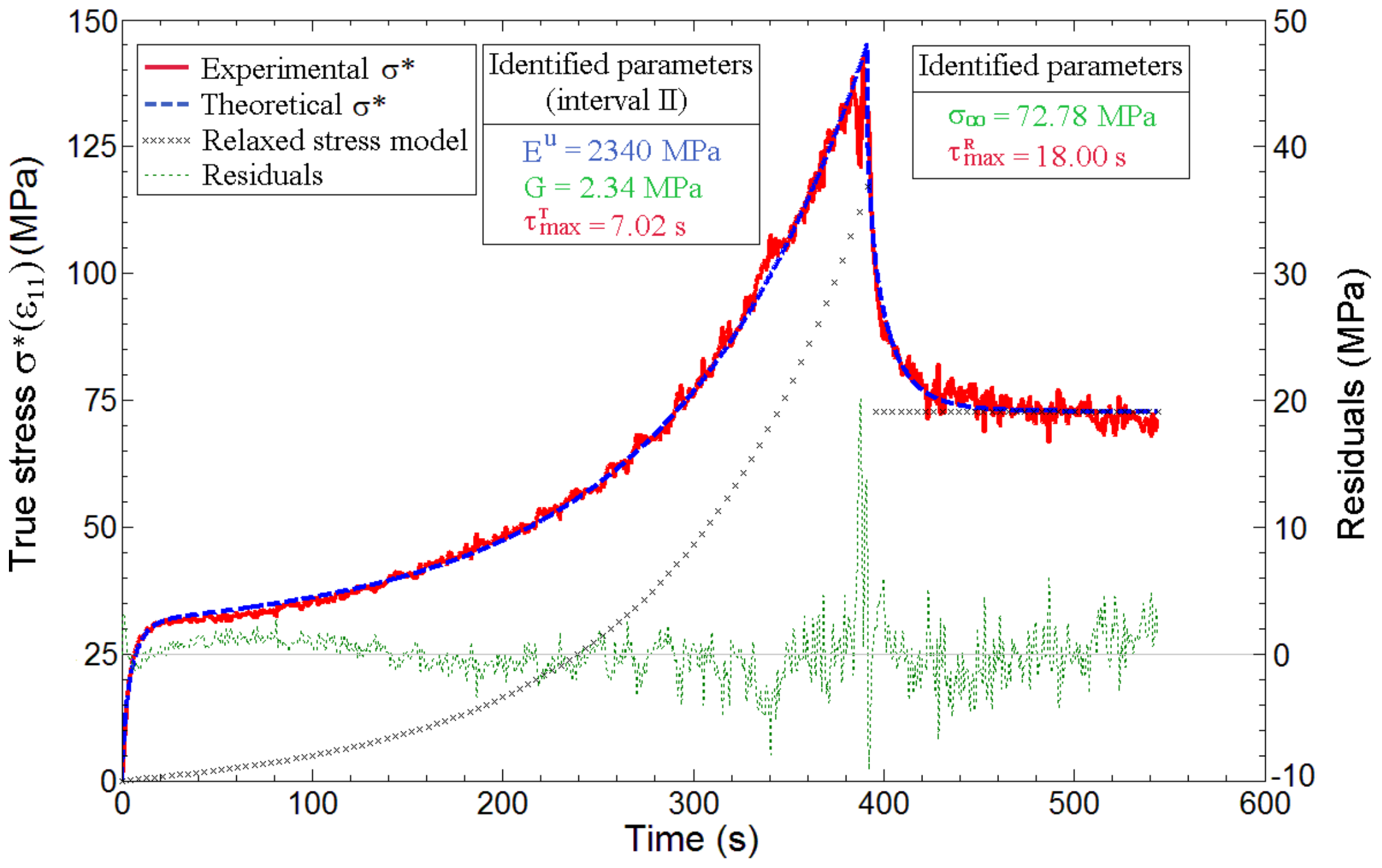

Figure 10: Tensile test followed by a relaxation. Experimental data, fitted curve and residuals as a function of time: identified parameters $E^u, G, \tau_{max}{}^T, \sigma_\infty$ and $\tau_{max}{}^R$ - specimen B - Strain rate of $5\times10^{-3} s^{-1}$ : 400 s corresponds to a true strain of 2).

### 4.2 Sensitivity Analysis

Fig. 11 plots the sensitivity coefficients (Eq. (24)) of the model parameters and table 2 presents the results of the stochastic analysis.

| Interval I | $E^u$ | $\tau_{max}{}^T$ |
|---|---|---|
| $E^u$ | 2.15 % | -0.9667 |
| $\tau_{max}{}^T$ | -0.9667 | 3.61 % |

(a)

| Interval II | $E^u$ | G | $\tau_{max}{}^T$ |
|---|---|---|---|
| $E^u$ | 1.37 % | -0.18 | -0.9988 |
| G | -0.18 | 0.06 % | 0.147 |
| $\tau_{max}{}^T$ | -0.9988 | 0.147 | 1.38 % |

(b)

Table 2. Variance-covariance matrix of parameters $E^u$ and $\tau_{\max}{}^T$ on interval I (a) and $E^u$, G and $\tau_{\max}{}^T$ on interval II (b).

One can notice that both parameters $E^u$ and $\tau_{\max}{}^T$ are totally correlated on a large part of the curve (between $\varepsilon \approx 0.1$ and $\varepsilon \approx 0.8$). Obviously, if the identification was carried out only on the plateau, it would be completely impossible to get simultaneous information about $E^u$ and $\tau_{\max}{}^T$ as both elastic and viscoelastic regimes are over. Fig. 12 represents the normalized sensitivity to the elastic modulus $E^u$ versus the sensitivity to $\tau_{\max}{}^T$ in order to check the relation between both parameters all along the test. One can see that at small strains also ($\varepsilon \leq 0.13$), $E^u$ and $\tau_{\max}{}^T$ appear to be strongly correlated as the curve is not so far from the straight line with a zero intercept. The correlation coefficient between $E^u$ and $\tau_{\max}{}^T$ has a value very close to 1 (Tables 2a and 2b) as a result of the high degree of correlation between both parameters for most part of the curves (especially during the plastic regime). Nevertheless, the identification is possible as confirmed by the relative error calculated on both $E^u$ and $\tau_{\max}{}^T$ parameters. This is due to the sensitivities behavior around $\varepsilon_{11} = 0.13$ where a singularity marks the transition between the different regimes (insert of figure 12). One can also notice that parameter $G$ can be estimated with a very low confidence bound (error of 0.06 %) as a consequence of its very weak (inexistent) correlation with the other parameters ($\rho_{12} = -0.18$ and $\rho_{23} = 0.14$). The robustness of the identification has also been checked by changing the initial values of the optimization algorithm. If two parameters were correlated, then the estimated values would have changed for each different run. Converging to the identified parameter vector whatever its initialization, is a good verification that the estimation is well-posed.

Regarding the use of interval I to perform the identification on parameters $E^u$ and $\tau_{\max}{}^T$ only, the sensitivity analysis helps to its definition. For strains below 0.13, the sensitivity to $G$ is only 5% of the maximum sensitivities of $E^u$ and $\tau_{\max}{}^T$ and therefore, this parameter can be omitted. It is interesting to note by comparing table 2a and 2b that the variances on the parameters $E^u$ and $\tau_{\max}{}^T$ is lowered by a factor of 2 when considering interval II instead of interval I. Therefore it is clear that using the whole curve is preferable.

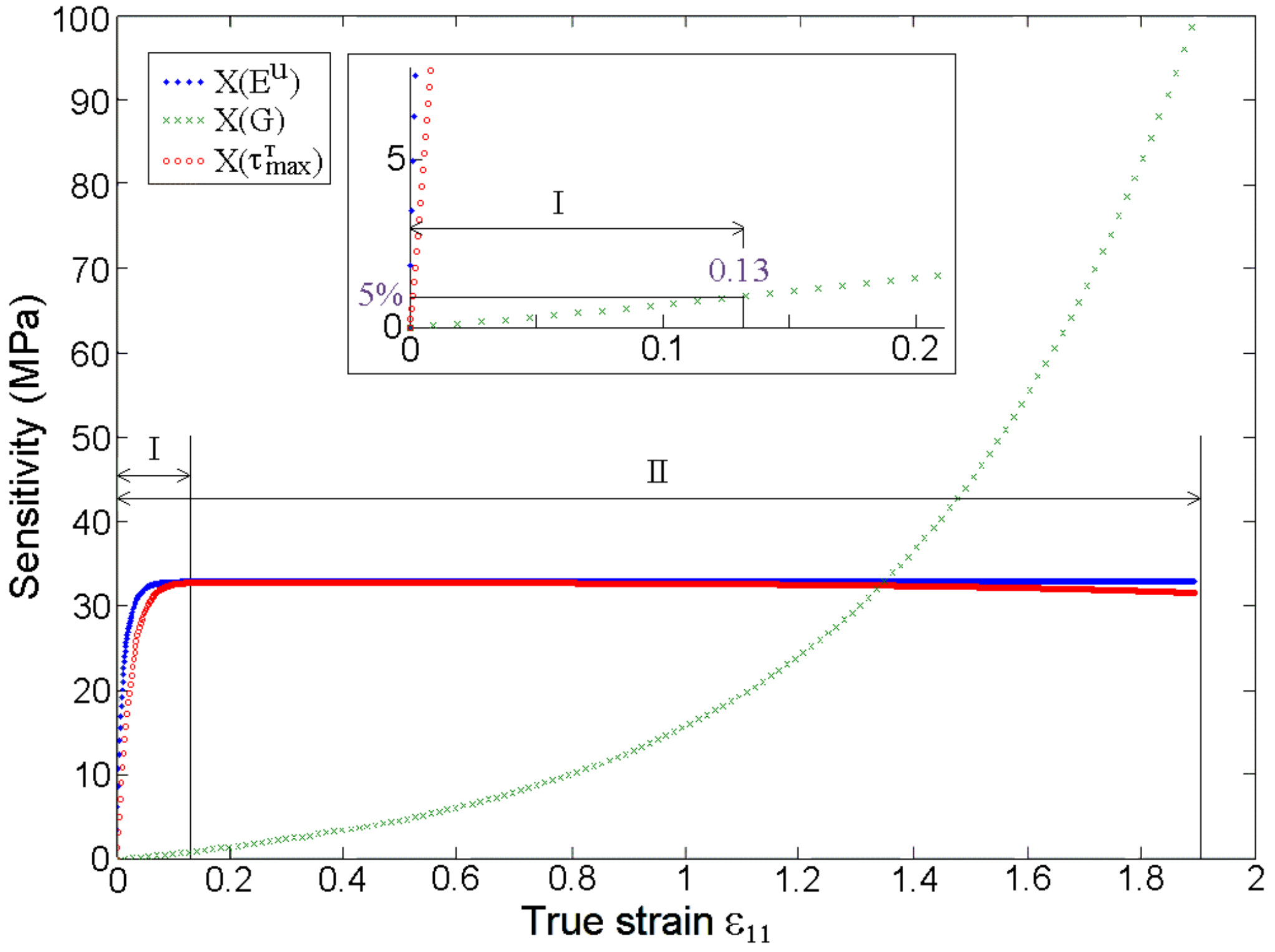

Figure 11: Sensitivities to parameters $E^u$, G and $\tau_{max}{}^T$ on interval II.

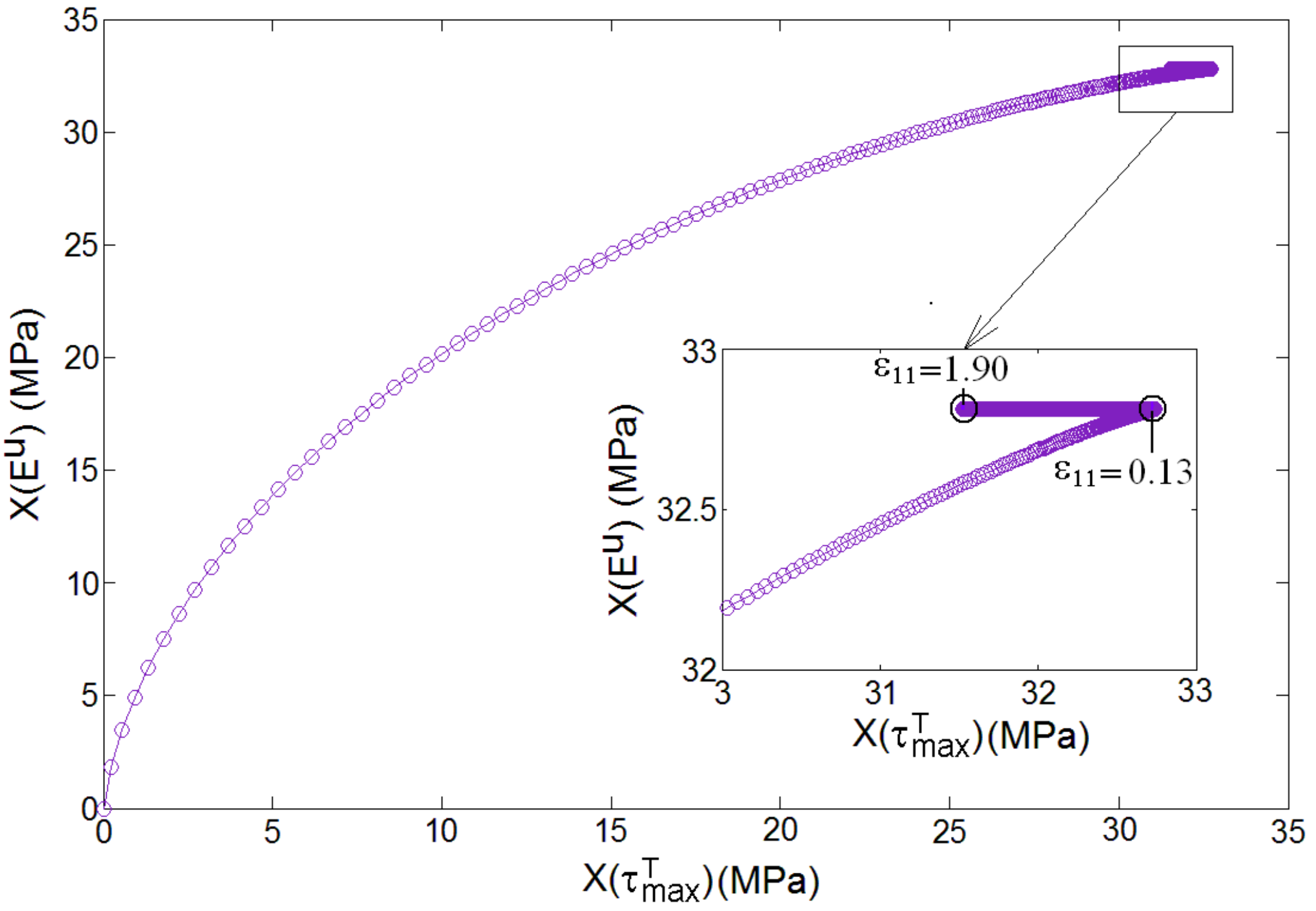

Figure 12: Sensitivity $X(E^u)$ as a function of $X(\tau_{max}{}^T)$.

## 5. Estimated parameters and discussion

Table 3 gives the values estimated for specimen A on both interval I ($E^u, \tau_{max}{}^T$) and interval II ($E^u, \tau_{max}{}^T, G$).

| Strain rate | $2.5\times10^{-3}s^{-1}$ | $5\times10^{-3}s^{-1}$ | | | $10^{-2}s^{-1}$ | |
|---|---|---|---|---|---|---|
| Experiment | 1 | 1 | 2 | 3 | 1 | 2 |
| Eu (MPa) **Interval I** | 2787 | 2741 | 2826 | 2917 | 2851 | 2841 |
| Eu (MPa) Interval II | 2940 | 2854 | 2725 | 2856 | 2819 | 2726 |
| $\tau_{max}{}^T$ (s) **Interval I** | 11.40 | 6.18 | 6.07 | 5.78 | 3.13 | 3.20 |
| $\tau_{max}{}^T$ (s) Interval II | 10.64 | 5.79 | 6.29 | 5.95 | 3.15 | 3.36 |
| G (MPa) | 2.31 | 2.32 | 2.28 | 2.32 | 2.49 | 2.44 |

Table 3. Identified values of parameters $E^u, \tau_{max}{}^T$ (Identification interval I) and $E^u, \tau_{max}{}^T, G$ (Identification interval II) - Specimen $A_{//}$.

### 5.1. Identification of the elastic modulus $E^u$

As can be seen from Table 3, the elastic moduli identified for sample A, specimen $A_{//}$, lie in the range from 2700 to 2950 MPa for all experiments (including those at different strain rates). This represents a variation of $\pm 4\%$ around the central value which corresponds to the order of magnitude of the estimated variances yielded by the stochastic analysis (Table 2b). The uncertainty on these measurements has been (over)estimated by changing the value of the identified $E^u$ within the $\pm\Delta E^u$ interval in order that the direct model produces theoretical curves that flank the experimental ones. This allows to take into account the bias effects evidenced earlier by the residuals plotting. For specimen $A_{//}$, at a strain rate of $5\times10^{-3}s^{-1}$, this uncertainty is of $\pm 180\,MPa$ which corresponds to a variation of $\pm 6.5\%$ around the nominal value. The results for all tested specimen are : $2830^{\pm 180}\ MPa$, $2940^{\pm 200}\ MPa$, $2400^{\pm 260}\ MPa$, and $2270^{\pm 150}\ MPa$ for specimens $A_{//}$, $A_c$, $A_\perp$ and B respectively. These values have been reported in Table 4 and will be commented later.

In order to check the validity of the results obtained by this MBM approach, two independent direct measurements of the elastic modulus of the material have been carried out.

**Ultrasound measurements:** The first one rests upon a pulse-echo technique and an ultrasonic device (Optel OPBOX system). Direct measurements are made of the velocity of a sonic wave in the longitudinal and transverse directions of the sample. Two transducers enable sonic excitation (range 0.5 to 10 MHz) thanks to piezoelectric components which apply shear and compression stresses on a ceramic membrane. The transducers collect echoes of the sonic wave after reflection on the backside of the sample. Then, the measured signal enables calculating the average period $T$ between two successive echoes. For a semicrystalline polymer, three echoes can usually be obtained (millimetric thicknesses). Longitudinal and transverse speeds of the wave are derived using the following formula:

$$V = \frac{2e}{T} \tag{38}$$

where $2e$ is twice the thickness of the sample.

Once the density $\rho$ of the material, the transverse and the longitudinal velocities ($V_T$, $V_L$) are known, the elastic modulus can be calculated as follows:

$$E_{PE}^{u} = \frac{\rho\, V_T^{\,2} \left(3\, V_L^{\,2} - 4\, V_T^{\,2}\right)}{V_L^{\,2} - V_T^{\,2}} \tag{39}$$

The OPBOX system offers different ways to calculate the velocities.

- A single identified echo can be located by the experimentalist. The system, through identification of the envelope curve of the oscillating signal, determines the time-of-flight (maximum of the echo) and calculate the velocity according to (38).
- Multiple echoes can be used (successive or not). The time interval separating two echoes is determined thanks to idealized envelope curves accounting for attenuation of the signal.
- The experimentalist can measure directly the time interval using cursors located on well-identified extrema of repeated pattern (echoes). This method was found less accurate.

These three possibilities allow us to check the confidence that can be placed in the measurements (through reproducibility and consistence) and to calculate a measurement uncertainty.

For specimens $A_{//}$ and $A_c$, values of respectively $2780^{\pm 40}\ MPa$ and $2800^{\pm 230}\ MPa$ have been found at room temperature (Table 4). (Of course, there is no possibility of measuring a Young modulus in a specific direction, hence $E^u_{A_{//}} = E^u_{A_\perp}$, and $E^u_{A_{//}}$ must be considered as the modulus of the 6mm thick material). Likewise, the measurements on specimen B gave a value of $2220^{\pm 230}\ MPa$ with measurements of the longitudinal sound velocity very conform to those correlated to density in Piché (1984) and both longitudinal and transversal velocities reported by Legros et al. (1999).

**Nanoindentation and CSM method:** The second direct measurement is based on nanoindentation tests where the Young's modulus (and hardness) are deduced using the continuous stiffness method (CSM) (Olivier et al., 1992; Le Rouzic et al., 2009). Experiments have been carried out on our samples with CSM at L.M.A. Femto-ST (Laboratoire de Mécanique Appliquée, Besançon-France) on a Nanonindenter II$^S$ using a pyramidal Berkovich probe tip. In this technique, the indenter vibrates at a frequency of $45Hz$ for an amplitude $\Delta h_0$ of *1-2 nm* during the indenter penetration over a depth $h$ varying from *0.5* to $6\ \mu m$. This process generates elastic and plastic deformations which results in a print conforming to the shape of the indenter. In the CSM method, the small harmonic load oscillation $F = F_0 \exp(i\omega t)$ is superimposed to the static one and, knowing the deformation response of the material $\Delta h = \Delta h_0 \exp(i\omega t)\exp(i\phi)$, a complex modulus can be identified, the real component being the apparent instantaneous (storage) modulus $E'$.

It is measured through direct calculations according to Le Rouzic et al., 2009.

$$E' = S_m \frac{\sqrt{\pi}}{2\eta\sqrt{A}} \cos(\phi) \quad (40)$$

where $\phi$ is the phase lag due to viscous dissipation, $S_m = F_0 / \Delta h_0$ the modulated stiffness, and $A$, the projected area of the elastic contact. For a Berkovich indenter, constant $\eta = 1.034$ in eq. (40) and $A$ is given by the following formula

$$A \approx 24.56\ h_c^2 \quad (41)$$

with $h_c = h - \varepsilon F / S_m$ and where $F$ is the force recorded along the displacement of the indenter, and $\varepsilon = 0.72$ for a conical indenter.

The true modulus of the specimen can be derived from the apparent measurement by taking into account the stiffness of the indenter through relation:

$$\frac{1}{E'} = \frac{1-\nu^2}{E} + \frac{1-\nu_i^2}{E_i} \quad (42)$$

where ($\nu$, $E$) and ($\nu_i$, $E_i$) stand for the Poisson ratio and elastic modulus respectively for the material and the indenter. But for polymer application, $E_i >> E$ $(E_i = 1040 GPa)$ and eq. (42) is reduced to:

$$E' = \frac{E}{1-\nu^2} \quad (43)$$

Combining equations (40), (41) and (43) leads to:

$$E^{u}_{Nano} = E = \frac{S_m}{2h_c\sqrt{\frac{24.56}{\pi}\frac{\eta}{1-\nu^2}}} \quad (44)$$

Figure 13 gives the results in terms of modulus $E^{u}_{Nano} = E$ as a function of the penetration depth. The measurements are repeated over several indents (between N=15 and N=20 for each sample) with a $50\mu m$ distance between them. Only measurements obtained for penetration depths ranging from $3-6\mu m$ and above have been considered in order to avoid uncertainties in the detection of the surface sample and specimen roughness effects (Qasmi and Delobelle, 2006). Although, the samples were polished using a $2\mu m$ alumina based polishing paste. Compared to measurements made on the initial rough surface, this treatment has been shown to have a great impact on the reproducibility and the confidence bounds, which have been reduced by a factor of nearly 20. The measurements made on polished specimens are routinely within a $\pm 50MPa$ confidence interval. According to the imposed displacement path of the indenter, the experiments correspond to a strain rate of about $\dot{\varepsilon} = 2\times 10^{-2}\, s^{-1}$ $\left(\varepsilon = 1/h \times dh/dt\right)$ which is exactly the order of magnitude of the tensile tests. The difference lies in the volume of matter tested ($\mu m$ scale). Taking a value of 0.5 for the Poisson coefficient (as was made for tensile tests), the following values have been found for the instantaneous modulus:

Specimen B: $2180^{\pm 110} MPa$, Specimen $A_{//}$ : $2380^{\pm 40} MPa$, Specimen $A_{\perp}$ : $2320^{\pm 40} MPa$, specimen $A_c$ : $2660^{\pm 40} < E_{A_c} < 2790\, MPa$ (Table 4).

The technique also offers the possibility to measure the stress at 3.3% of true strain through hardness measurements. Hb values in the range $80-90\, MPa$ have been obtained leading to $\sigma_{0.03} = Hb/3 = 27-30MPa$, a value very close to the one that can be read on Fig 6.

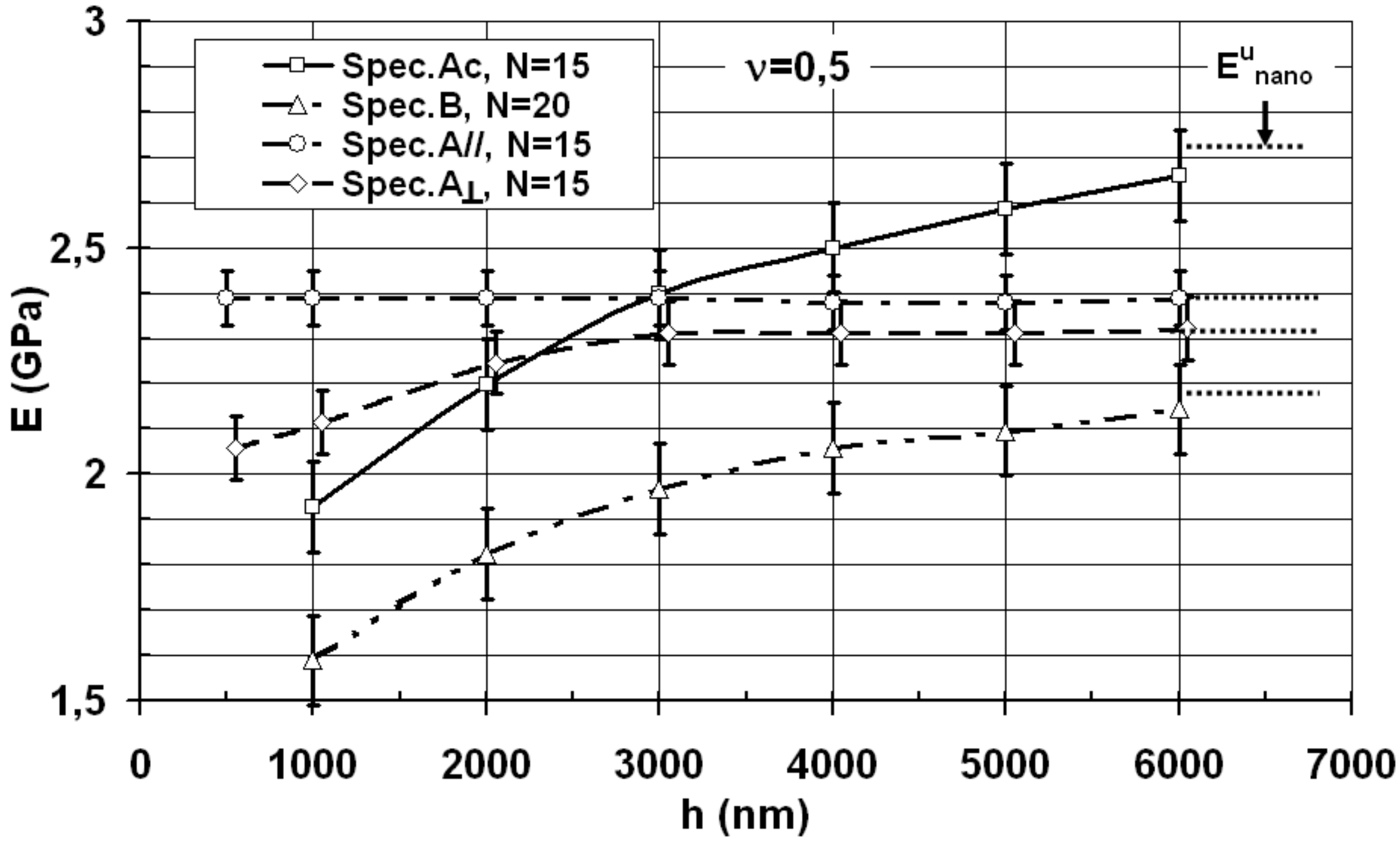

Figure 13: Modulus identified from Nanoindentation tests versus penetration depth (all specimen, N repeated experiments).

**Comparison between MBM, PE and Nanonindentation CSM methods:** Table 4 gathers the values of the elastic modulus obtained for all HDPE Natural 500 specimens A ( $A_{//}$ , $A_{\perp}$ and $A_c$ ) and B by the three metrological techniques (indices MBM, PE and Nano for respectively the **M**odel **B**ased **M**etrology applied to tensile curves, the **P**ulse-**E**cho and **Nano**nindentation technique). The data obtained by applying standards on tensile curves are also given. It is evident from these latter values that standards are very far from producing (i) reliable data and (ii) physically founded estimations. The discrepancy can be as large as 300% and the values depend on the applied displacement rate (Table 1)! Nevertheless, using tensile curves to identify precisely the elastic modulus of polymers is still possible on the condition that a correct MBM is used. This can be asserted from the comparison made in Table 4 with the values yielded by the two other physical techniques based on direct measurements.

| Specimen | | Manufac-turer | ISO 527-1 $d\varepsilon_N/dt$ = 5 mm/min | ASTM D638 $d\varepsilon_N/dt$ = 5 mm/min | Present Model-Based parameter estimation $E^u_{MBM}$ | Ultrasonic technique $E^u_{PE}$ | Nanoindentation $E^u_{Nano}$ |
|---|---|---|---|---|---|---|---|
| A | $A_{//}$ | 1200 | 1138 | 1141 | 2830 | 2780 (-2%) | 2380 (-16%) |
| | $A_c$ | | - | - | 2940 | 2800 (-5%) | 2730 (-8%) |
| | $A_\perp$ | | - | - | 2400 | 2780 (= $E^u_{A_{//}}$ ) | 2320 (-3%) |
| B | | | 772 | 763 | 2270 | 2220 (-2%) | 2180 (-5%) |

Table 4. Comparison between elastic modulus values given by standards (from tensile curve), the present model-based parameter estimation, the pulse-echo technique and the nanoindentation technique, for all samples A ( $A_{//}$ , $A_\perp$ , $A_c$ ) and B (in MPa). Discrepancy percentages in brackets calculated with reference to $E^u_{MBM}$ .

In view of the uncertainty intervals given above (not reported in Table 4), the following comments can be made. For all specimens, the three techniques give very close results. The agreement is very good between MBM and PE values but local differences can occur with the Nanoindentation test. These differences must be analyzed carefully. For specimen $A_{//}$ , the value given by the Nano technique is 16% lower than for the MBM and PE techniques. This can be explained as the Nanotechnique probes a small volume just behind the surface. Therefore the identified value may be sensitive to some slight skin effect which may be present on the 6 mm extruded sample ( $A_{//}$ ). On the contrary, both the MBM and PE methods probe the entire volume. As specimen B was manufactured directly with a 4 mm thickness, this effect is not visible (within the confidence intervals of both three measurements). Now looking at specimen $A_c$ (3mm core of specimen $A_{//}$ ), this effect should disappear. Higher values $E^u_{Nano}$ are indeed obtained and both three techniques give more convergent results. Despite the measurement uncertainty, it seems that a general trend can be observed: for all three techniques, the measured elastic modulus is higher for the core specimen. This may prove that (i) core samples are slightly more rigid (small differences in microstructure) and (ii), the MBM technique is sensitive to such small variation.

Regarding now the incidence of these results on the material property, it is clear that also this commercial HDPE is produced under the same reference, it can exhibit large changes in its elastic behaviour (around $600\,MPa$). The increase in elasticity for samples $A_{//}$ , $A_c$ compared to sample B indicates an oriented texture due to the extrusion process. The presence of such anisotropy was confirmed by a positive anisotropy index measured for the undeformed HDPE specimens $A_{//}$ , thanks to X-ray microtomography (Blaise et al., 2010). This result is then conform to those given in Legros et al. (1999) where un-oriented and oriented HDPE are investigated through Pulse-Echo technique. Oriented HDPE is shown to produce higher longitudinal and shear wave signal and hence higher elastic modulus.

### 5.2. Identification of the maximal relaxation time $\tau_{max}^T$

Table 5 gathers the estimated maximal relaxation times $\tau^T_{\max}$ for all specimens A and B and the three different strain rates. Also shown is the corresponding Deborah number (ratio between the time constant $t_{mat}$ characterizing intrinsic ‘fluidity’ of the material, and the time scale of the experiment $t_{\exp}$ or of the observer). The fluidity of the material is inversely proportional to the Deborah number. In the present case, its maximal value can be calculated by the following formula:

$$De_{\max} = \frac{t_{mat}}{t_{\exp}} = \tau_{\max}\ \dot{\varepsilon} \tag{45}$$

where $\tau_{max}$, the maximum relaxation time of the spectrum, is used for the material characteristic time and $1/\dot{\varepsilon}$ for the experimental characteristic time of excitation.

| Specimen | Strain rate ($s^{-1}$) | $\tau_{max}^{T}$ (s) | $De_{max}$ | G (MPa) |
|---|---|---|---|---|
| $A_{//}$ | $2.5.10^{-3}$ | 11.02 | 0.0276 | 2.31 |
| | $5.10^{-3}$ | 6.01 | 0.0301 | 2.31 |
| | $10^{-2}$ | 3.21 | 0.0321 | 2.46 |
| B | $5.10^{-3}$ | 7.16 | 0.0358 | 2.30 |
| | $10^{-2}$ | 3.79 | 0.0379 | 2.39 |

Table 5. Average estimated maximal relaxation time $\tau_{\max}{}^{T}$, corresponding Deborah number $De_{\max}$ and hardening modulus $G$ for specimens $A_{//}$ and B and different strain rates.

It can be pointed out that the Deborah numbers found for the range of investigated strain rates remain very close. They are of the order of 0.03 which means that the time response of the material is two decades below the time scales imposed by the experiment. This argues in favor of a similar scenario for the succession of internal mechanisms taking place at the microstructure levels. A slight increase of this value is maybe observed when strain rate increases. The difference between the highest and smallest values is about 15% for specimen A which is a little bit higher than the expected variance on this parameter. The maximal relaxation time identified by the model seems then to express a slight sensitivity of the material with respect to the applied strain rate (as confirmed by the tensile curves of Fig. 3). This may confirm the results of an in-situ microstructural investigation based on light scattering and infrared imaging during cold drawing (Baravian et al., 2008), where kinetic effects appear starting from a strain rate of $10^{-2} s^{-1}$. The increase in Deborah number agrees well with the observation that plasticity of the material is enhanced when the strain rates are lower. Finally it is interesting to recall that a simple linear viscoelastic assumption was made for the spectrum specification. The relatively constant Deborah numbers found whatever the applied strain rates (and for a same quality of the identification process, with same behavior of the residuals) validates this hypothesis.

### 5.3. Identification of the hardening modulus *G*

Concerning the value of the hardening modulus $G$ describing the behavior of HDPE at high strains, the identified values for all strain rates are also very close (discrepancies are less than 5% as shown by Table 5). Those identified values for $G$ are close to the ones found in the experimental results of uniaxial tensile tests reported by Haward on HDPE (Haward, 1993; Haward, 2007) and the value found by Bartczak and Kozanecki (2005) for linear polyethylene in plane-strain compression. The estimated hardening modulus appears to be the same for both specimens A and B. A slight increase of the modulus is observed for the highest strain rate ($10^{-2}$ $s^{-1}$) but this value might also be slightly biased as shown by the quite important signed character of the residuals. This may indicate that at large strain rates, the Haward-Thackray approach used as relaxed state in the model may become inappropriate.

### 5.4. Conclusion about the identification of the parameters

Parameters $E^u, {\tau_{\max}}^T$ and G of the reduced model presented in section 3 have been shown to be easily identifiable thanks to an appropriate sensitivity analysis. For each tested strain rate, the same elastic modulus has been identified as expected by the instantaneous character of this physical parameter. The hardening modulus and Deborah number based on the maximal relaxation time of the spectrum also are found to be nearly constant. A slight effect due to strain rate is obvious, but remains limited. It is accompanied by more pronounced wavy residuals, indicating that some bias may exist between the idealized model and its experimental realization when the strain rate increases. Thanks to the application of this physical model, one can retrieve much precise information regarding the behavior of the polymer. Finally, the results found by this model for the Young modulus $E^u$ have been corroborated by two other independent and direct physical measurements probing the material at high frequencies (MHz) or at small spatial scales (μm). This is a clear evidence of the consistence of the approach with the underlying physics because this result is the corollary of an indentified maximum relaxation time of the order of a few seconds. As a spectrum of 6 decades is considered in our approach to get the convergence of the model, this means that relaxation times as small as 1 μs are necessary to allows for a good description with the model and a determination of a ‘truly’ instantaneous modulus. If one considered that the pulse-echo technique relies on an excitation at 5 MHz, this means that the time scale required by the model is in perfect agreement with this technique which in turn give a support to our identification of the elastic modulus. Note that for this technique, the spatial size of the representative elementary volume is of the order of a few micrometers, which means also that small time scale events are probed.

## 6. Conclusion

An inverse approach has been presented for the estimation of the parameters of a rheological law of SCP's in the case of a tensile test. Both ideas of using an appropriate reduced model in relationship to sensitivity analysis principles are shown to be the key for pertinent measurements through fitting procedures. Illustration is given for the characterization of different HDPE specimens through a tensile test. The identified instantaneous Young modulus is shown to be twice the value determined by contemporary recommended standards thus showing that progress has to be made in this metrological field to improve the characterization of polymers which exhibit elastoviscoplastic behavior. It is unsensitive to the strain rates, as expected from such a property. Finally it is shown to be very close to the determinations made according to two different physical techniques probing the sample at very low time scales (Pulse-echo technique) or very small length scales (Modulated Nanoindentation test). Other conclusions regarding the viscoelastic and hardening behaviors also suggest the relevancy of the approach.